\documentclass[article,authoryear,longauth]{aa}
\usepackage{graphicx}
\usepackage{txfonts}
\usepackage{xcolor}
\usepackage{amsmath}
\usepackage{subcaption}

\usepackage[hidelinks]{hyperref}
\hypersetup{
  colorlinks,
  linkcolor={red},
  citecolor={blue!80!black},
  urlcolor={blue!80!black}
}
\usepackage{color}
\usepackage{siunitx}

\newcommand{\ticA}{{\rm TOI-5153}}
\newcommand{\ticBtop}{{\rm NGTS-20 (=TOI-5152)}}
\newcommand{\ticB}{{\rm NGTS-20}}

\begin{document}

\title{Two long-period transiting exoplanets on eccentric orbits: NGTS-20\,b (TOI-5152\,b) and TOI-5153\,b}

\titlerunning{Two long-period transiting exoplanets on eccentric orbits}

\author{
  S.~Ulmer-Moll \inst{\ref{geneva}} \and
  M.~Lendl \inst{\ref{geneva}} \and
  S.~Gill \inst{\ref{warwick}, \ref{ inst3 }} \and
  S.~Villanueva \inst{\ref{ inst7 }} \and
  M.J.~Hobson \inst{\ref{ inst8 }} \and
  F.~Bouchy \inst{\ref{geneva}} \and
  R.~Brahm \inst{\ref{ inst9 }, \ref{ inst10 }} \and
  D.~Dragomir \inst{\ref{ inst11 }} \and
  N.~Grieves \inst{\ref{geneva}} \and
  C.~Mordasini \inst{\ref{ inst26 }} \and
  D.R.~Anderson \inst{\ref{warwick}, \ref{ inst3 }} \and
  J.~S.~Acton \inst{\ref{ inst12 }} \and
  D.~Bayliss \inst{\ref{warwick}, \ref{ inst3 }} \and
  A.~Bieryla \inst{\ref{ inst13 }} \and
  M.~R.~Burleigh \inst{\ref{ inst12 }} \and
  S.~L.~Casewell \inst{\ref{ inst12 }} \and
  G.~Chaverot \inst{\ref{geneva}} \and
  P.~Eigmüller \inst{\ref{ inst14 }} \and
  D.~Feliz \inst{\ref{ inst15 }, \ref{ inst16 }} \and
  S.~Gaudi \inst{\ref{ inst17 }} \and
  E.~Gillen \inst{\ref{ inst4 }, \ref{ inst5 }, \thanks{Winton Fellow}} \and
  M.~R.~Goad \inst{\ref{ inst12 }} \and
  A.~F.~Gupta \inst{\ref{ inst18 }, \ref{ inst19 }} \and
  M.~N.~Günther \inst{\ref{ inst20 }, \thanks{ESA Research Fellow}} \and
  B.~A.~Henderson \inst{\ref{ inst12 }} \and
  T.~Henning \inst{\ref{ inst8 }} \and
  J.~S.~Jenkins \inst{\ref{ inst22 }, \ref{ inst23 }} \and
  M.~Jones \inst{\ref{ inst24 }} \and
  A.~Jordán \inst{\ref{ inst9 }, \ref{ inst10 }} \and
  A.~Kendall \inst{\ref{ inst12 }} \and
  D.~W.~Latham \inst{\ref{ inst13 }} \and
  I.~Mireles \inst{\ref{ inst25 }} \and
  M.~Moyano \inst{\ref{ inst27 }} \and
  J.~Nadol \inst{\ref{ inst11 }} \and
  H.P.~Osborn \inst{\ref{ inst28 }, \ref{ inst7 }} \and
  J.~Pepper \inst{\ref{ inst29 }} \and
  M.~T.~Pinto \inst{\ref{ inst9 }} \and
  A.~Psaridi \inst{\ref{geneva}} \and
  D.~Queloz \inst{\ref{ inst5 }, \ref{ inst39 }} \and
  S.~Quinn \inst{\ref{ inst30 }} \and
  F.~Rojas \inst{\ref{ inst31 }, \ref{ inst32 }} \and
  P.~Sarkis \inst{\ref{ inst8 }} \and
  M.~Schlecker \inst{\ref{ inst33 }} \and
  R.~H.~Tilbrook \inst{\ref{ inst12 }} \and
  P.~Torres \inst{\ref{ inst31 }, \ref{ inst32 }} \and
  T.~Trifonov \inst{\ref{ inst8 }, \ref{ inst38 }} \and
  S.~Udry \inst{\ref{geneva}} \and
  J.~I.~Vines \inst{\ref{ inst35 }} \and
  R.~West \inst{\ref{warwick}} \and
  P.~Wheatley \inst{\ref{warwick}, \ref{ inst3 }} \and
  X.~Yao \inst{\ref{ inst29 }, \ref{ inst36 }} \and
  Y.~Zhao \inst{\ref{geneva}} \and
  G.~Zhou \inst{\ref{ inst37 }} }

\authorrunning{Ulmer-Moll et al.}

\institute{
  Observatoire de Genève, Université de Genève, Chemin Pegasi, 51, 1290 Versoix, Switzerland \label{geneva} \and
  Department of Physics, University of Warwick, Gibbet Hill Road, Coventry CV4 7AL, UK \label{warwick} \and
  Centre for Exoplanets and Habitability, University of Warwick, Gibbet Hill Road, Coventry CV4 7AL, UK \label{ inst3 } \and
  Department of Physics and Kavli Institute for Astrophysics and Space Research, Massachusetts Institute of Technology, Cambridge, MA 02139, USA \label{ inst7 } \and
  Max-Planck-Institut für Astronomie, Königstuhl 17, D-69117 Heidelberg, Germany \label{ inst8 } \and
  Facultad de Ingeniería y Ciencias, Universidad Adolfo Ibáñez, Av. Diagonal las Torres 2640, Peñalolén, Santiago, Chile \label{ inst9 } \and
  Millennium Institute for Astrophysics, Chile \label{ inst10 } \and
  Department of Physics and Astronomy, University of New Mexico, 1919 Lomas Blvd NE Albuquerque, NM, 87131, USA \label{ inst11 } \and
  Physikalisches Institut, University of Bern, Gesellschaftsstrasse 6, 3012 Bern, Switzerland \label{ inst26 } \and
  School of Physics and Astronomy, University of Leicester, Leicester LE1 7RH, UK \label{ inst12 } \and
  Center for Astrophysics $\|$ Harvard \& Smithsonian, 60 Garden Street, Cambridge, MA 02138, USA \label{ inst13 } \and
  Institute of Planetary Research, German Aerospace Center, Rutherfordstrasse 2., 12489 Berlin, Germany \label{ inst14 } \and
  Fisk University, Nashville TN, USA \label{ inst15 } \and
  Vanderbilt University, Nashville TN, USA \label{ inst16 } \and
  Department of Astronomy, Ohio State University, 140 W. 18th Ave., Columbus, OH, 43210, USA \label{ inst17 } \and
  Astronomy Unit, Queen Mary University of London, Mile End Road, London E1 4NS, UK \label{ inst4 } \and
  Astrophysics Group, Cavendish Laboratory, J.J. Thomson Avenue, Cambridge CB3 0HE, UK \label{ inst5 } \and
  Department of Astronomy \& Astrophysics, 525 Davey Laboratory, The Pennsylvania State University, University Park, PA, 16802, USA \label{ inst18 } \and
  Center for Exoplanets and Habitable Worlds, 525 Davey Laboratory, The Pennsylvania State University, University Park, PA, 16802, USA \label{ inst19 } \and
  European Space Agency (ESA), European Space Research and Technology Centre (ESTEC), Keplerlaan 1, 2201 AZ Noordwijk, Netherlands \label{ inst20 } \and
  Núcleo de Astronomía, Facultad de Ingeniería y Ciencias, Universidad Diego Portales, Av. Ejército 441, Santiago, Chile \label{ inst22 } \and
  Centro de Astrofísica y Tecnologías Afines (CATA), Casilla 36-D, Santiago, Chile \label{ inst23 } \and
  European Southern Observatory, Alonso de Córdova 3107, Vitacura, Casilla 19001, Santiago, Chile \label{ inst24 } \and
  Department of Physics and Astronomy, University of New Mexico, 210 Yale Blvd NE Albuquerque, NM, 87106, USA \label{ inst25 } \and
  Instituto de Astronomía, Universidad Católica del Norte,Angamos 0610, 1270709, Antofagasta, Chile \label{ inst27 } \and
  Center for Space and Habitability, University of Bern, Gesellschaftsstrasse 6, 3012, Bern, Switzerland \label{ inst28 } \and
  Department of Physics, Lehigh University, 16 Memorial Drive East, Bethlehem, PA 18015, USA \label{ inst29 } \and
  ETH Zurich, Department of Physics, Wolfgang-Pauli-Strasse 27, 8093 Zurich, Switzerland \label{ inst39 } \and
  Center for Astrophysics, Harvard \& Smithsonian, 60 Garden Street, Cambridge, MA 02138, USA \label{ inst30 } \and
  Instituto de Astrofísica, Pontificia Universidad Católica de Chile, Avda. Vicuña Mackenna 4860, Macul, Santiago, Chile \label{ inst31 } \and
  Millennium Institute of Astrophysics, Nuncio Monsenor Sotero Sanz 100, Of. 104, Providencia, Santiago, Chile \label{ inst32 } \and
  Department of Astronomy/Steward Observatory, The University of Arizona, 933 North Cherry Avenue, Tucson, AZ 85721, USA \label{ inst33 } \and
  Department of Astronomy, Sofia University "St Kliment Ohridski", 5 James Bourchier Blvd, BG-1164 Sofia, Bulgaria \label{ inst38 } \and
  Departamento de Astronomía, Universidad de Chile, Casilla 36-D, Santiago, Chile \label{ inst35 } \and
  Shanghai Astronomical Observatory, Chinese Academy of Sciences, 80 Nandan Road, Shanghai 200030, People's Republic of China \label{ inst36 } \and
  Centre for Astrophysics, University of Southern Queensland, West Street, Toowoomba, QLD 4350 Australia \label{ inst37 }
}

\date{Received 18 March 2022 / Accepted 6 July 2022}

\abstract
    {Long-period transiting planets provide the opportunity to better understand the formation and evolution of planetary systems.
      Their atmospheric properties remain largely unaltered by tidal or radiative effects of the host star, and their orbital arrangement reflects a different,
      and less extreme, migrational history compared to close-in objects.
      The sample of long-period exoplanets with well determined masses and radii is still limited,
      but a growing number of long-period objects reveal themselves in the Transiting Exoplanet Survey Satellite (TESS) data.}
    {Our goal is to vet and confirm single transit planet candidates detected in the TESS space-based photometric data
      through spectroscopic and photometric follow up observations with ground-based instruments.}
    {We work with high-resolution spectrographs to confirm the planetary nature of the transiting candidates and measure their masses.
      We use the Next Generation Transit Survey (NGTS) to photometrically monitor the candidates in order to observe additional transits.
      Using a joint modeling of the light curves and radial velocities, we compute the orbital parameters of the system and are able to precisely measure the mass and radius of the transiting planets.
    }
    { We report the discovery of two massive, warm Jupiter-size planets,
      one orbiting the F8-type star \ticA\ and the other orbiting the G1-type star \ticBtop.
      From our spectroscopic analysis, both stars are metal-rich with a metallicity of 0.12 and 0.15, respectively.
      Only \ticA\ presents a second transit in the TESS extended mission data, however
      NGTS observed \ticB\ as part of its mono-transit follow-up program and detected two additional transits.
      Follow-up high-resolution spectroscopic observations were carried out with CORALIE, CHIRON, FEROS, and HARPS.
      \ticA\ hosts a 20.33 day period planet with a planetary mass of
      $\rm 3.26 ^{+0.18} _{-0.17}$ Jupiter masses ($M_{J}$), a radius of $\rm 1.06 ^{+0.04} _{-0.04}\,R_{J}$, and an orbital eccentricity of $0.091^{+0.024}_{-0.026}$.
      \ticB\ b is a $\rm 2.98 ^{+0.16} _{-0.15}\,M_{J}$ planet with a radius of $1.07 ^{+0.04} _{-0.04}\,R_{J}$
      on an eccentric ($0.432 ^{+0.023} _{-0.023}$) orbit with an orbital period of 54.19 days.
      Both planets are metal-enriched and their heavy element content is in line with
        the previously reported mass-metallicity relation for gas giants.}
    {Both warm Jupiters orbit moderately bright host stars making these objects valuable targets
      for follow-up studies of the planetary atmosphere and measurement of the spin-orbit angle of the system.}

    \keywords{Planetary systems --
      Planets and satellites: detection --
      Planets and satellites: individual: \ticB\ / TOI-5152 --
      Planets and satellites: individual: \ticA\ --
      Planets and satellites: gaseous planets --
      methods: data analysis
    }

  \maketitle

%

  \section{Introduction}
%

  The majority of known transiting gas giants are hot Jupiters; they orbit their host star with periods of less than ten days. The extended atmospheres of hot Jupiters are ideal for atmospheric studies and understanding their composition (e.g. \citealt{madhusudhan_2014a,sing_2016,wyttenbach_2017,showman_2020,baxter_2021}), however the original properties of these systems are not often preserved (e.g. \citealt{albrecht_2012}).
  The orbital parameters of an exoplanet are the result of its origin and formation but in the case of close-in planets such as hot Jupiters, the proximity of the host star leads to additional mechanisms such as tidal interactions (e.g. \citealt{valsecchi_2015}) which disturb the initial orbital parameters. Longer period transiting planets, such as warm Jupiters, are less affected by their host star and their orbital elements do retain a record of their formation and migrational history.
  
  Warm Jupiters are exoplanets orbiting their host star with periods typically defined between 10 and 200 days. Similarly to hot Jupiters, if warm Jupiters are not formed in-situ, they require migration mechanisms which are able to bring them from several au to a fraction of an au from their host star \citep{dawson_2018}. Part of the warm Jupiter population is located in the period valley, a region of the parameter space between 10 and 100 days where gas giants are less frequent (\citealt{udry_2003,wittenmyer_2010}). The occurrence rates of gas giants can also suggest which type of formation and evolution different gas giants undergo. One of the distinctive features of the warm Jupiter population is its wide range of eccentricities. The formation and evolution processes leading to the diversity of orbital arrangements seen in systems hosting warm Jupiters remain to be fully understood.

  Several explanations have been put forward to explain the wide eccentricity distribution of warm Jupiters.
  The eccentricity distribution of warm Jupiters is composed of two groups, a low eccentricity component and a higher eccentricity one (e.g. \citealt{petrovich_2016}). Disk migration (e.g. \citealt{goldreich_1980, baruteau_2014}) is able to explain the lower eccentricity component and can reproduce the period distribution of gas giants under certain disk properties \citep{coleman_2016}. However, disk migration and ensuing planet-planet scattering do not create enough warm Jupiters with high eccentricities \citep{petrovich_2014}. In high-eccentricity migration scenarios (e.g. \citealt{rasio_1996,fabrycky_2007}), warm Jupiters are precursors of hot Jupiters which we are observing in the midst of inward migration. High-eccentricity migration produces a satisfying number of warm Jupiters at high eccentricities but not enough low eccentricity warm Jupiters. Besides, high-eccentricity migration under-produces warm Jupiters \citep{wu_2011}, while disk migration produces a number of warm Jupiters in agreement with the estimate of the period valley. Thus it is thought that a combination of these mechanisms could explain the observed population.

  The composition of gas giants depends on where they are formed in the protoplanetary disk, on which timescales, but also on the composition of the disk. Measuring precise masses and radii of giant planets enables estimations of their bulk metallicity, which in turn constrains the formation and evolution models. Solar system giants are metal-enriched compared to the solar metallicity (e.g. \citealt{wong_2004}) and core-accretion models are able to match their heavy metal enhancement (e.g. \citealt{alibert_2005}). For exoplanets, \citet{thorngren_2016b} showed that the planet metal enrichment ($Z_{planet}/Z_{star}$) is anti-correlated with the planetary mass while the mass of heavy elements is positively correlated with planetary mass. Comparing the results of different planetary synthesis models with the bulk metallicity and atmospheric composition of exoplanets leads to constraints on the processes driving core formation and envelope enrichment (e.g. \citealt{mordasini_2014,mordasini_2016a}). However, the sample of warm Jupiters with precise mass and radius measurements is scarce. And the correlation between planet metal-enrichment and planetary mass still needs further investigation, for example in terms of detailed host star abundances \citep{teske_2019} and in terms of its dependence on orbital properties of the planetary systems \citep{dalba_2022}.

  Detecting transiting warm Jupiters is challenging, especially for ground-based surveys, as partial or full transit events visible from a given site are rare. However, the space-based photometric mission Transiting Exoplanet Survey Satellite (TESS; \citealt{ricker_2015a}) is opening a new window to detect this type of exoplanet orbiting bright and well characterized stars. TESS observes each field almost continuously for 27 days, but a significant fraction of the sky is covered for longer periods of time when fields overlap. As such, TESS monitors some parts of the sky for up to one year in regions called continuous viewing zones. For fields observed continuously for 27 days, we expect most warm Jupiters to show only a single transit. Simulations by \cite{cooke_2018} and \cite{villanueva_2019a} predicted that between 500 and 1000 single transit events would be found in the first two years of the TESS data. Following these results, several pipelines have been developed to search for these events (e.g. \citealt{gill_2020,montalto_2020a}). These dedicated searches provide valuable candidates in addition to the ones announced by the TESS Objects of Interest (TOIs; \citealt{guerrero_2021}).

  Single transit candidates have several observational disadvantages, namely their orbital period is largely unconstrained. An estimate of the orbital period can be determined based on stellar and transit parameters \citep{osborn_2016}. With the re-observation of previous TESS sectors through the extended mission, new single transit candidates are detected and several of the known candidates show a second transit, constraining the possible planetary periods to a discrete set of values. These candidates require spectroscopic vetting as we expect a false positive rate of about 50\% \citep{santerne_2016}. Hence a by-product of the search for warm Jupiters is the detection and characterization of long-period low-mass eclipsing binaries (e.g. \citealt{lendl_2020,gill_2022}). Single and duo transit candidates are challenging systems to follow up and characterize but they can reveal warm planets which are highly valuable systems (e.g. \citealt{osborn_2022,schanche_2022}). Long-period transiting giants are the missing link between hot Jupiters and the solar system giants and they offer precious information for understanding the physics of their atmosphere, their formation, migration and evolution history.

  This paper describes the discovery and characterization of two new warm Jupiters. The observations are detailed in Section~\ref{observations}, the derivation of stellar parameters and the methods used to analyze photometric and radial velocity data are explained in Section~\ref{methods}. The results are presented in Section~\ref{results} and discussed in Section~\ref{discussion}. We summarize our findings in Section~\ref{conclusion}.

\section{Observations}
\label{observations}

The discovery photometry was collected with the space-based mission TESS (Section~\ref{sec:tess}) and follow-up observations were carried out from the ground with the photometric facility NGTS (Section~\ref{sec:ngts}), and the high-resolution spectrographs CORALIE, FEROS, CHIRON, HARPS, and TRES (Section~\ref{sec:coralie}, ~\ref{sec:feros}, ~\ref{sec:chiron}, ~\ref{sec:harps}, and \ref{sec:tres}). The presence of nearby stars was checked with speckle imaging (Section~\ref{sec:nessi}).

\subsection{TESS photometry}
\label{sec:tess}

\ticA\ and \ticB\ were observed by the TESS satellite during its primary and extended mission. Both targets showed a single transit event in sectors 4 and 6 respectively. As several teams set out to search for and vet this type of events, these stars were selected as single transit candidates by the TSTPC (Tess Single Transit Planetary Candidate) group, and later announced as CTOIs (Community Tess Object of Interest) by J. Steuer and the WINE team.
\ticA\ was observed at a 30 min cadence in sector 6 (2018-12-11 to 2019-01-07) and at a 2 min cadence in sector 33 (2020-12-17 to 2021-01-13).
\ticB\ was observed at a 30 min cadence in sector 4 from 2018-10-18 to 2018-11-15 and at a 2 min cadence in sector 31 from 2020-10-21 to 2020-11-19.
Both stars were observed at a 2 min cadence in the extended mission, thanks to the approved Guest Investigator Program G03188 led by S. Villanueva.
However, only \ticA\ showed a second single transit in sector 33 and no event was detected in sector 31 for \ticB.
Both primary transits passed the vetting stage where we checked for asteroid crossing and centroid shifts (indicating of a background eclipsing binary).  

\begin{figure*}
  \includegraphics[width=0.95\hsize]{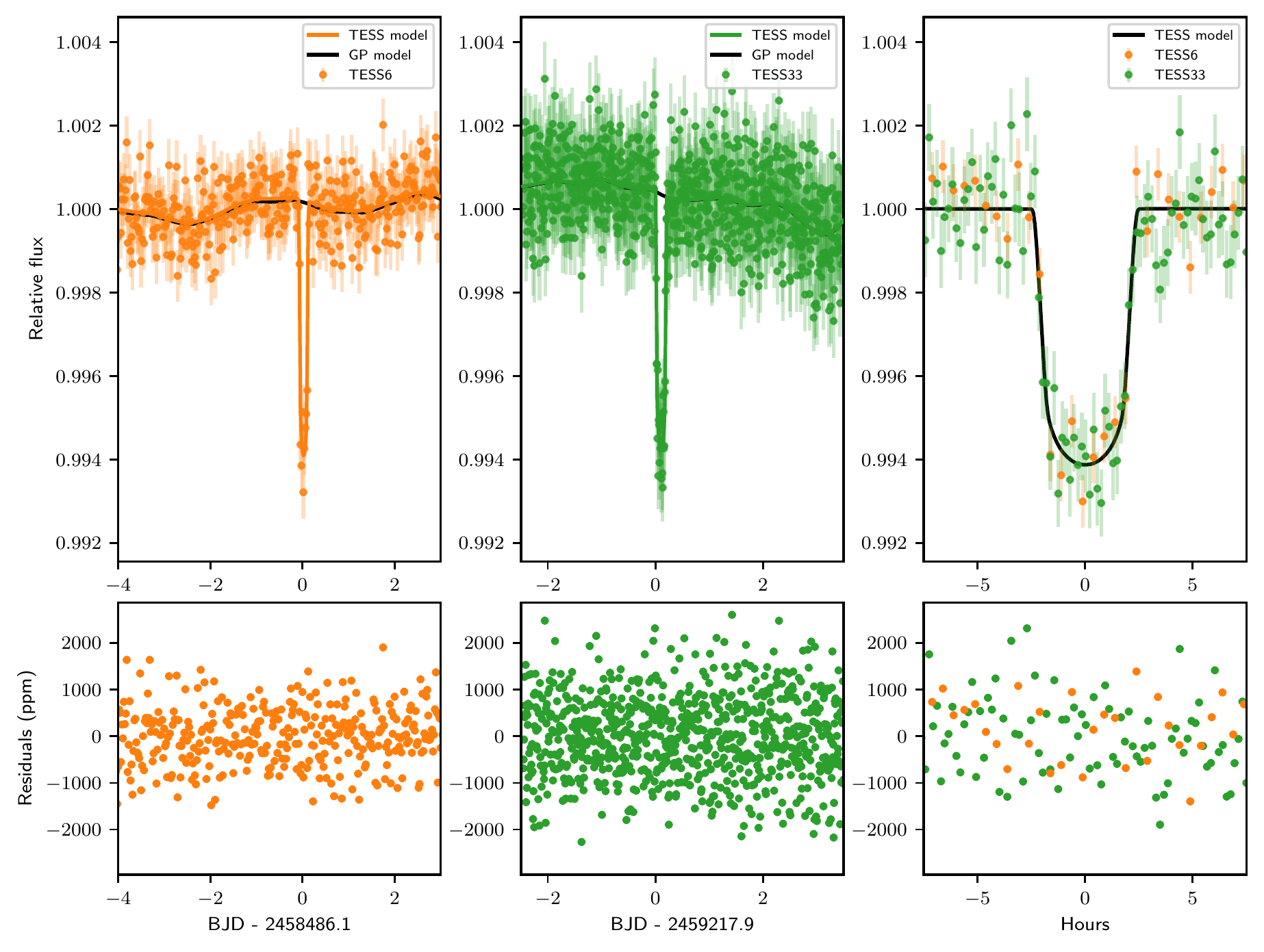}
  \caption{Top: Photometric observations of \ticA\ from TESS sector 6 at 30 min cadence (orange dots, left panel) and sector 33 at 10 min cadence (green dots, middle panel) with full median models (orange and green lines), and Gaussian process models (black line). Right side panel shows the detrended and phase folded data from both sectors (orange and green dots) with the phase folded transit model in black. Bottom: Each panel show the residuals in parts per million between the full model and the respective light curve.}
  \label{fig:phase-folded_lc_1240}
\end{figure*}

\begin{figure*}
  \includegraphics[width=\hsize]{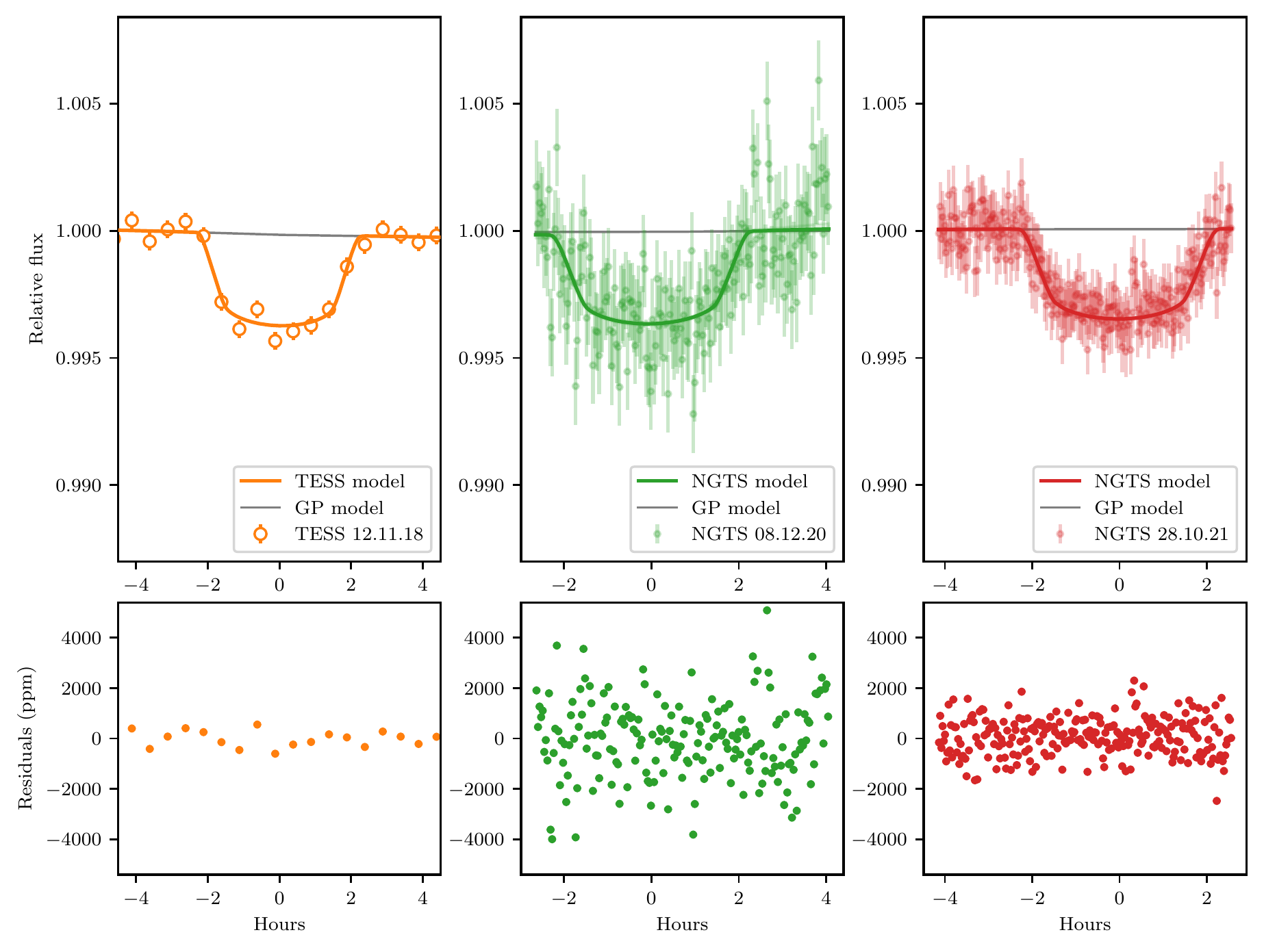}
  \caption{Top: Photometric observations of \ticB\ from TESS sector 4 at 30 min cadence (left panel) and NGTS binned at 2 min cadence (middle and right panels). In each panel the data is shown are colored dots (orange, green, and red), the full model is represented with a line of the same color and the Gaussian process model as a grey line. Bottom: Each panel show the residuals in parts per million between the full model and the respective light curve.}
  \label{fig:phase-folded_lc_2575}
\end{figure*}

Light curves of the primary mission were extracted with the Quick Look Pipeline (QLP, \citealt{huang_2020,huang_2020a}).
The light curves from the extended mission were obtained through the data reduction done at the Science Processing Operation Center (SPOC, \citealt{jenkins_2016}). We use the Simple Aperture Photometry (SAP) fluxes and their corresponding errors for our analysis.
The light curves are presented in Figures~\ref{fig:phase-folded_lc_1240} \& \ref{fig:phase-folded_lc_2575}.
We generated target pixel files with \texttt{tpfplotter} \citep{aller_2020} and checked the presence of contaminant sources, down to a magnitude difference of 6, within the aperture used to extract the light curves.
One star (TIC\,124029687) falls into the aperture around \ticA. We estimate the dilution by comparing the fluxes measured in the Gaia passband RP as this filter matches the TESS passband. \ticA\ has a mean RP flux of 273878 $\pm$ 52 electrons per second ($\rm e^{-1}s^{-1}$) and TIC\,124029687 has a flux of 2208\,$\pm$\,8\,$\rm e^{-1}s^{-1}$. The dilution is equal to 0.8$\pm$0.003\% (Equation 6 from \citealt{espinoza_2019a}). We included a dilution factor for the light curve modeling and we chose a Normal prior informed by the dilution estimated with Gaia photometry.
The aperture of \ticB\ is not contaminated by neighboring stars, hence we chose to fix the dilution factor to 1 (no dilution) for the modeling of the light curve.

\subsection{NGTS photometry}
\label{sec:ngts}

\ticB\ was monitored from the ground by the Next Generation Transit Survey (NGTS). NGTS is an automated array of twelve 20\,cm telescopes installed at ESO's Paranal Observatory, Chile \citep{wheatley_2018}. NGTS uses a custom filter which spans 520 to 890\,nm. Starting on the night of September 29th, 2020, the target was observed in the blind survey mode (every possible night) with one 20\,cm telescope. The data was acquired with an exposure time of 10 seconds and cadence of 13 seconds. Data reduction is performed with standard aperture photometry and an automatic transit search is done using template matching \citep{gill_2020a}. One full transit was observed on the night of December 8th, 2020. After this second transit event, the possible periods for this candidate correspond to a set of period aliases. The target was only observed on nights when a transit of a period alias was expected. A third transit was observed on the night of October 28th, 2021. Six NGTS cameras were used with the same exposure time and cadence as during the blind search survey mode.
The NGTS light curves are presented in Figure~\ref{fig:phase-folded_lc_2575}.

\begin{table}
  \caption{Radial velocities of \ticA.}
  \label{table:table_rvs_1240}
  \begin{tabular}{l l l l}
    \hline
    \hline
    \noalign{\smallskip}
    Time             & RV                   & RV error                  & Instrument\\
    BJD              & [$\rm km\,s^{-1}$]     & [$\rm km\,s^{-1}$]          & \\
    \hline
    \noalign{\smallskip}
    2459192.69733	& -35.24751	& 0.06266	& CORALIE\\
    2459201.70551	& -35.56908	& 0.06674	& CORALIE\\
    2459214.55618	& -35.30501	& 0.08927	& CORALIE\\
    ...&&&\\
    2459504.78439       & -35.4622      & 0.0156        & HARPS\\
    2459505.85753       & -35.5273      & 0.0213        & HARPS\\
    2459506.76824       & -35.5145      & 0.0162        & FEROS\\
    
    \hline
  \end{tabular}
  \tablefoot{Full table is available at CDS.}
\end{table}

\begin{table}
  \caption{Radial velocities of \ticB.}
  \label{table:table_rvs_2575}
  \begin{tabular}{l l l l}
    \hline
    \hline
    \noalign{\smallskip}
    Time             & RV                   & RV error                 & Instrument\\
    BJD              & [$\rm km\,s^{-1}$]     & [$\rm km\,s^{-1}$]         & \\
    \hline
    \noalign{\smallskip}
    2458738.86129    & 0.0778               & 0.0282                   & CHIRON\\
    2458748.83030    & 0.1598               & 0.0273                   & CHIRON\\
    2458804.74919    &  12.67750            & 0.05916                  & CORALIE\\
    ...&&&\\
    2459504.73569 & 12.60444             & 0.03076                  & CORALIE\\
    2459514.61823 & 12.66770             & 0.02817                  & CORALIE\\
    2459528.58470 & 12.43836             & 0.03839                  &  CORALIE\\

    \hline
  \end{tabular}
  \tablefoot{Full table is available at CDS.}
\end{table}

\subsection{CORALIE spectroscopy}
\label{sec:coralie}

Spectroscopic vetting and radial velocity follow-up was carried out with the CORALIE spectrograph \citep{queloz_2001a}. CORALIE is a fiber-fed spectrograph installed at the Nasmyth focus of the Swiss 1.2m Euler telescope (La Silla, Chile). CORALIE has a spectral resolution of 60\,000 and observes with a 3 pixel sampling per resolution element. Fiber injection is done with two fibers: a first fiber is used to observe the target and a second fiber can collect light from a Fabry-Pérot etalon or the sky to allow for simultaneous wavelength calibration or background subtraction. 

\ticA\ and \ticB\ are part of an on-going CORALIE survey which aims to confirm TESS single transit candidates and characterize the properties of the systems. Spectroscopic vetting is first done by taking two spectra of the target about one week apart. These two observations are used to rule out eclipsing binary scenarios. Then each target is monitored with an average of one point per week, and the sampling is adapted to maximize the phase coverage of the orbit once a periodic signal is detected.

Stellar radial velocities are measured with the cross-correlation technique: the stellar spectrum is cross-correlated with a mask close to the stellar type of the host star to obtain a cross-correlation function (CCF, e.g. \citealt{pepe_2002a}). In addition to the radial velocity and its associated error, full width half maximum, contrast, and bisector inverse slope (BIS) are some of the parameters derived from the CCF. These parameters have been shown to be reliable tracers of the radial velocity noise induced by stellar activity and they can be used to detrend the data in some cases (e.g. \citealt{melo_2007}).

\begin{figure}
  \includegraphics[width=\hsize]{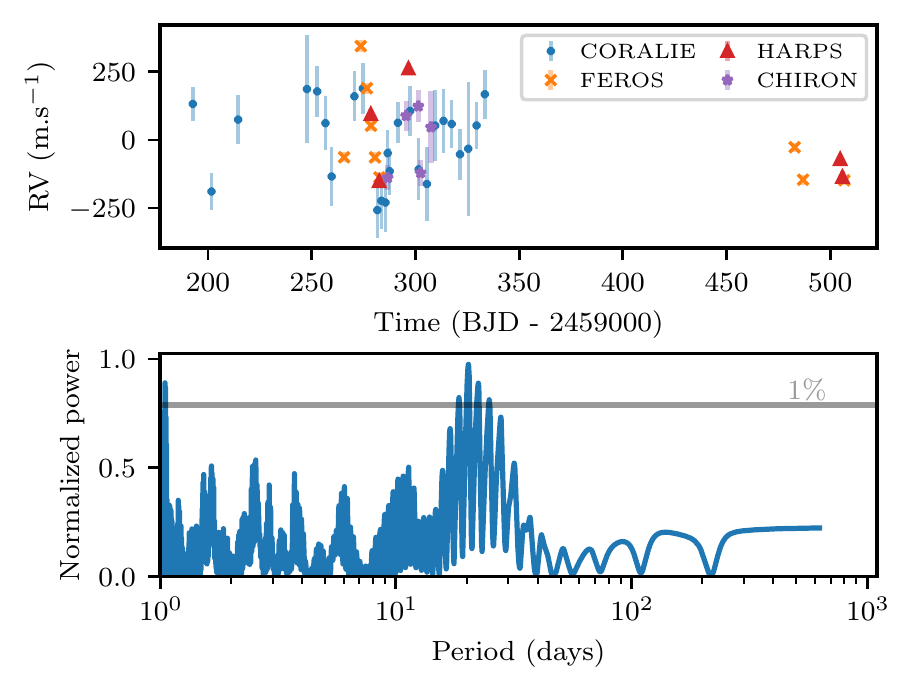}
  \caption{Top: Time series of the radial velocities from CORALIE (blue dots), FEROS (orange crosses), HARPS (red triangles), and CHIRON (purple stars) for \ticA. Bottom: Generalized Lomb-Scargle periodogram of the radial velocities. The highest peak corresponds to a period of about 20.1 days.}
  \label{fig:rv_periodo_1240}
\end{figure}

\begin{figure}
  \includegraphics[width=\hsize]{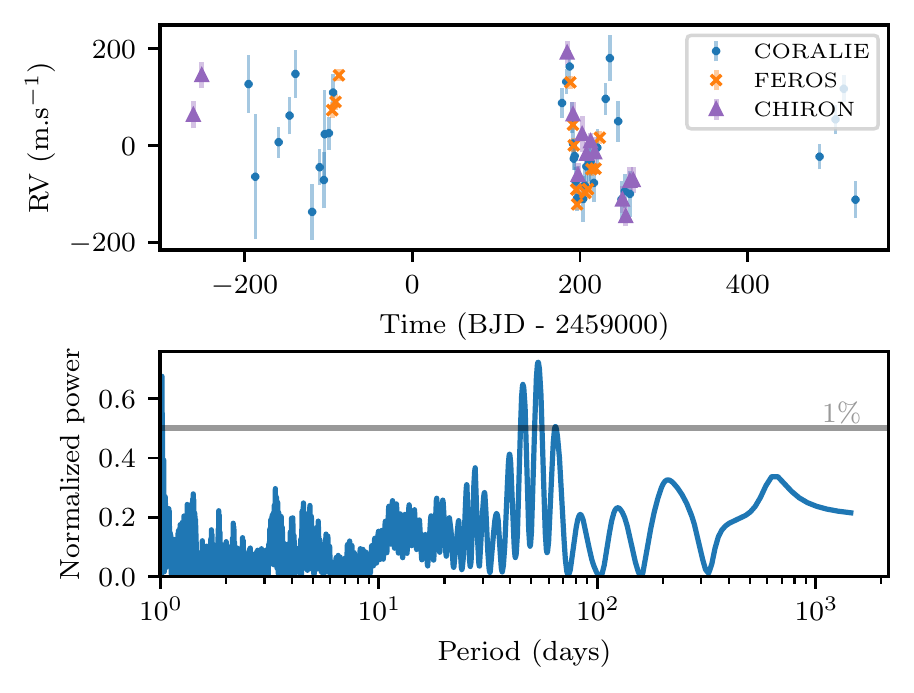}
  \caption{Top: Time series of the radial velocities from CORALIE (blue dots), FEROS (orange crosses), and CHIRON (purple triangles) for \ticB. Bottom: Generalized Lomb-Scargle periodogram of the radial velocities. The highest peak corresponds to a period of about 53.5 days.}
  \label{fig:rv_periodo_2575}
\end{figure}

We collected 25 radial velocity measurements of \ticA\ (from 2020-12-09 to 2021-04-28) and 39 measurements of \ticB\ (from 2019-11-17 to 2021-11-10) with an exposure time varying between 900 and 1800s. The spectra of \ticA\ have an average signal-to-noise ratio of 14 and the ones of \ticB\ have an average signal-to-noise of 23.
The observations for both targets are detailed in Tables~\ref{table:table_rvs_1240} \& \ref{table:table_rvs_2575} and plotted in Figures~\ref{fig:rv_periodo_1240} \& \ref{fig:rv_periodo_2575}. CORALIE spectra were also used to derive stellar parameters and the analysis is detailed in Section~\ref{stellar-analysis}.

\subsection{FEROS spectroscopy}
\label{sec:feros}

FEROS is a high-resolution spectrograph installed at the 2.2m telescope in La Silla, Chile. FEROS has a spectral resolution of 48\,000 with 3 pixel sampling. Two fibers are available to observe the target and simultaneously record the spectrum of Th-Ar lamp to allow a precise wavelength calibration \citep{kaufer_1999}.

A total of nine FEROS spectra were collected for \ticA\ between 2021-02-20 and 2021-10-09 under the program number 0106.A-9014(A) (PI: Sarkis). The exposure times vary between 900 and 1200 seconds depending on the weather conditions.
For \ticB, 13 spectra were recorded under the program number 0104.A-9007(A) (PI: Sarkis) between 2020-02-25 and 2021-01-09. Exposure times vary between 900s and 1350s depending on weather conditions and lead to a signal-to-noise ratio ranging from 70 to 123.

The data reduction was performed with the CERES pipeline \citet{brahm_2017}, and the radial velocities were extracted using the cross-correlation technique. Radial velocity measurements are detailed in Tables~\ref{table:table_rvs_1240} \& \ref{table:table_rvs_2575} and plotted in Figures~\ref{fig:rv_periodo_1240} \& \ref{fig:rv_periodo_2575}.

\subsection{CHIRON spectroscopy}
\label{sec:chiron}

\ticA\ was observed with the CHIRON spectrograph \citep{tokovinin_2013} of the 1.5 m Smarts telescope
located at Cerro Tololo International Observatory (CTIO) in Chile. CHIRON is a fiber-fed spectrograph with a spectral resolution of 80\,000
when used with the image slicer mode. Five spectra were taken from 2021-03-13 to 2021-04-03.
These data were acquired through a monitoring program from Sam Quinn and reduced with a
least-squares deconvolution method \citep{donati_1997} leading to an average radial velocity precision of \SI{67}{\meter\per\second}.

\ticB\ was monitored with the CHIRON spectrograph and observations took place between 2019-09-12 and 2021-02-18.
A total of 13 radial velocity measurements were obtained with an exposure time of 1800s and a nominal signal-to-noise ratio of 30.
The data was obtained through two different observing programs and the data reduction was done following the procedures described in \citet{wang_2019a} and \citet{jones_2019}, and wavelength calibration is done with Th-Ar lamp exposures taken before and after the science observation. The radial velocities were derived with the cross-correlation technique
and an average radial velocity precision of about \SI{27}{\meter\per\second} was reached. Radial velocity measurements are detailed in Tables~\ref{table:table_rvs_1240} \& \ref{table:table_rvs_2575} and plotted in Figures~\ref{fig:rv_periodo_1240} \& \ref{fig:rv_periodo_2575}.

\subsection{HARPS spectroscopy}
\label{sec:harps}

\ticA\ was also observed with the high-resolution spectrograph HARPS (\citealt{mayor_2003a}; $\rm R\sim115\,000$)
installed on the 3.6m telescope in La Silla, Chile. A total of three observations were
obtained under the program number 106.21ER.001 (PI: Brahm), between 2021-03-03 and 2021-03-23.
Observations were obtained with the high-accuracy mode and an exposure time set to 1200 seconds.
The data reduction was performed with the standard data reduction pipeline.
The radial velocities were extracted with the cross-correlation technique
using a G2 mask. We obtained an average signal-to-noise ratio of 25 at 550\,nm.
The observations are detailed in Table~\ref{table:table_rvs_1240} and plotted in Figure~\ref{fig:rv_periodo_1240}.

\subsection{TRES spectroscopy}
\label{sec:tres}
Two reconnaissance spectra of \ticA\ were obtained on February 10 and 19, 2022 using the Tillinghast Reflector Echelle Spectrograph (TRES; \citealt{furesz_2008}) located at the Fred Lawrence Whipple Observatory (FLWO) atop Mount Hopkins, Arizona, USA. TRES is a fiber-fed echelle spectrograph with a wavelength range of 390-910\,nm and a resolving power of 44\,000. The spectra were extracted as described in \citet{buchhave_2010}. We used the TRES spectra to derived stellar parameters for \ticA\ using the Stellar Parameter Classification (SPC) tool \citep{buchhave_2012,buchhave_2014}. SPC cross correlates an observed spectrum against a grid of synthetic spectra based on Kurucz atmospheric models \citep{kurucz_1992}. The stellar effective temperature is evaluated at $\rm 6190 \pm 53$\,K. The surface gravity is equal to $\rm 4.30 \pm 0.10\,cm\,s^{-2}$ and the stellar metallicity to $\rm 0.20 \pm 0.08$. These values are consistent within 1\,$\sigma$ with the adopted stellar parameters obtained from CORALIE spectra and described in Section~\ref{stellar-analysis}.

\begin{table*}
  \caption{Stellar properties and stellar parameters derived with the spectral synthesis method.}
  \label{table:stellar-params}
  \centering
  \begin{tabular}{l c c c}
    \hline
    \hline
    \noalign{\smallskip}
    & \ticA\ & \ticB\ & \\
    \hline
    \noalign{\smallskip}
    Other Names & & & \\
    \noalign{\smallskip}
    2MASS     & J06060966-1957118               & J03051020-2156011          & 2MASS\\
    Gaia      & 2942084865853011712             & 5078704372599743104        & Gaia\\
    TIC       & TIC 124029677                   & TIC 257527578              & TESS\\
    TOI       & TOI-5153                        & TOI-5152                   & TESS\\
    NGTS      & -                               & NGTS-20                    & NGTS\\
    \hline
    \noalign{\smallskip}
    Astrometric Properties & & & \\
    \noalign{\smallskip}
    R.A.                  & 06:06:09.68                & 03:05:10.23        & TIC\\
    Dec                   & -19:57:12.4                & -21:56:01.1        & TIC\\
    $\mu$R.A.($\rm mas\,yr^{-1}$)   & 8.745±0.019                & 20.077±0.017       & Gaia EDR3\\
    $\mu$Dec.($\rm mas\,yr^{-1}$)   & -34.563±0.021              & -0.873±0.017       & Gaia EDR3\\
    Parallax (mas)        & 2.563±0.023                & 2.731±0.018        & Gaia EDR3\\
    Distance (pc)         & 390.1±3.5                  & 366.2±2.4          & Gaia EDR3\\
    \hline
    \noalign{\smallskip}
    Photometric Properties & & & \\
    \noalign{\smallskip}
    V (mag)          & 11.93±0.15                   & 11.23±0.09          & Tycho\\
    B (mag)          & 12.46±0.19                   & 11.76±0.09          & Tycho\\
    G (mag)          & 11.5779±0.0004               & 11.0313±0.0004      & Gaia EDR3\\
    T (mag)          & 11.215±0.007                 & 10.6509±0.0078      & TESS\\
    J (mag)          & 10.681±0.024                 & 10.143±0.024        & 2MASS\\
    H (mag)          & 10.48±0.023                  & 9.879±0.025         & 2MASS\\
    Ks (mag)          & 10.408±0.019                 & 9.830±0.019         & 2MASS\\
    W1 (mag)         & 10.399±0.022                 & 9.800±0.023         & WISE\\
    W2 (mag)         & 10.412±0.02                  & 9.834±0.019         & WISE\\
    W3 (mag)         & 10.418±0.066                 & 9.775±0.036         & WISE\\
    W4 (mag)         & 8.595±0.361                  & 9.075±0.219         & WISE\\
    $\rm A_V$        & 0.13±0.04                    & 0.01±0.01           & Sec. 3.1\\
    \hline
    \noalign{\smallskip}
    Bulk Properties                        &                     &    & \\
    \noalign{\smallskip}
    $\rm T_{eff}$ (K)                       & 6300±80                       & 5980±80          & Sec. 3.1\\
    Spectral type                          & F8\,V                           & G1\,IV              & Sec. 3.1 \\
    log g ($\rm cm\,s^{-2}$)                 & 4.30±0.15                     & 3.8±0.2          & Sec. 3.1\\
    $\rm [Fe/H]$ (dex)                     & 0.12±0.08                     & 0.15±0.08          & Sec. 3.1 \\
    $\rm v.sini$ ($\rm km\,s^{-1}$)         & 10.1±1.0                      & 8.0±0.8            & Sec. 3.1\\
    $\rm log\,R'_{HK}$                      & -4.95±0.07                    & -5.01±0.05         & Sec. 3.1\\
    Age (Gyrs)                             & 5.4±1.0                       & 1.4 - 6.8          & Sec. 3.1\\
    Radius ($R_\odot$)                      & 1.40$\pm$0.04                 & 1.78$\pm$0.05      & Sec. 3.1\\
    Mass ($M_\odot$)                        & 1.24±0.07                     & 1.47$\pm$0.09      & Sec. 3.1\\
    \hline
  \end{tabular}
  \tablebib{2MASS \cite{skrutskie_2006}; GAIA EDR3 \cite{gaiacollaboration_2021}; Tycho \citep{hog_2000}; WISE \cite{wright_2010}.}
\end{table*}

\subsection{Speckle imaging}
\label{sec:nessi}

\begin{figure}
  \includegraphics[width=\hsize]{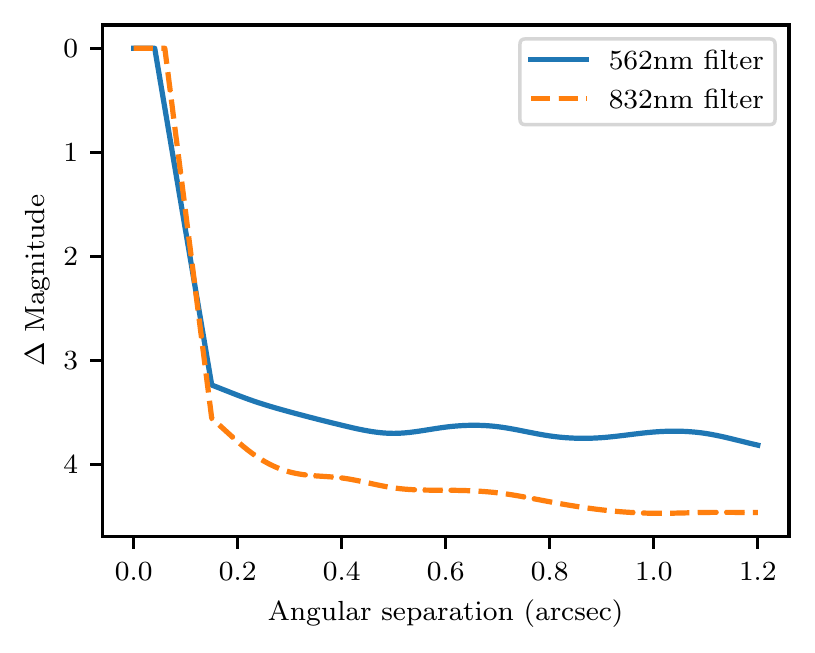}
  \includegraphics[width=\hsize]{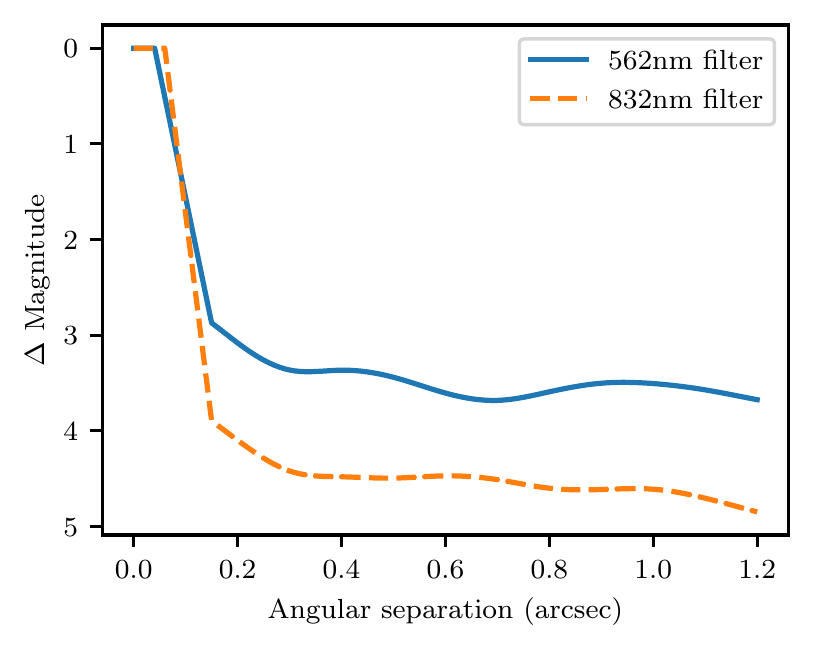}
  \caption{5-$\sigma$ background sensitivity curves derived from speckle images taken with NESSI for \ticA\ (top panel) and \ticB\ (bottom panel) showing no bright companion ($\Delta mag < 4$) from 0.2 to 1.2 arcsec.}
  \label{fig:nessi_1240}
\end{figure}

High-resolution speckle images were taken for both stars with the NN-Explore Exoplanet and Stellar Speckle Imager (NESSI: \citealt{scott_2018}). NESSI is installed on the 3.5m WYIN telescope located at the Kitt Peak National Observatory. \ticA\ was observed on the 17th of November, 2019 and \ticB\ was observed on the 18th of November, 2019. Both stars had images taken in the blue and red channels with two narrow band filters centered at 562\,nm and 832\,nm.
The reconstructed speckle images are produced following the procedures described in \cite{howell_2011}. The 5\,$\sigma$ background sensitivity limits are measured in the reconstructed images and are shown in Figure~\ref{fig:nessi_1240}. The NESSI data show no indication that either of the targets has close stellar companions.

\begin{figure}[!ht]
  \centering\includegraphics[width=6.2cm,angle=90,trim=130 50 50 100,clip]{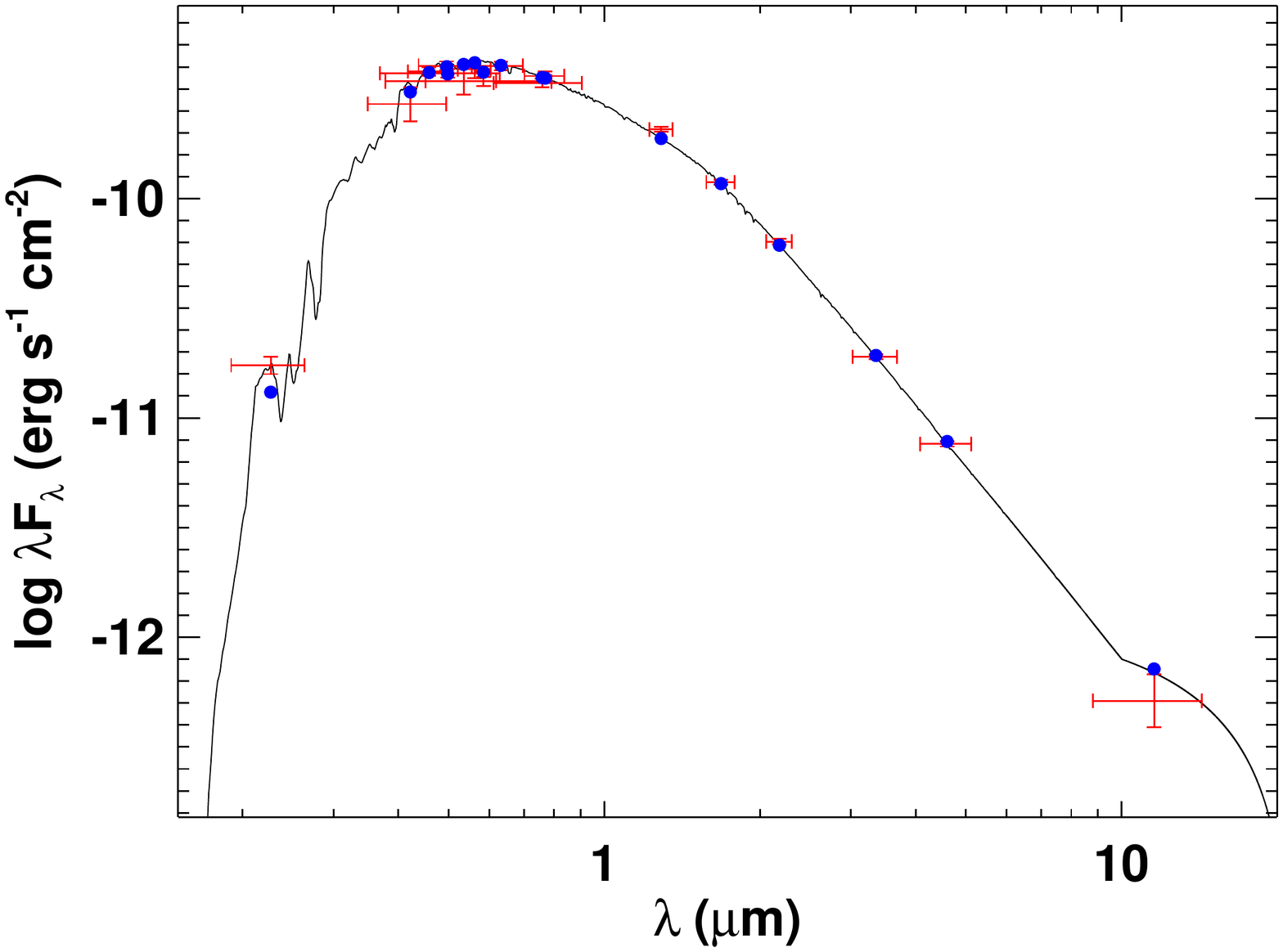}
  \centering\includegraphics[width=3.05cm,angle=90,trim=320 50 80 100,clip]{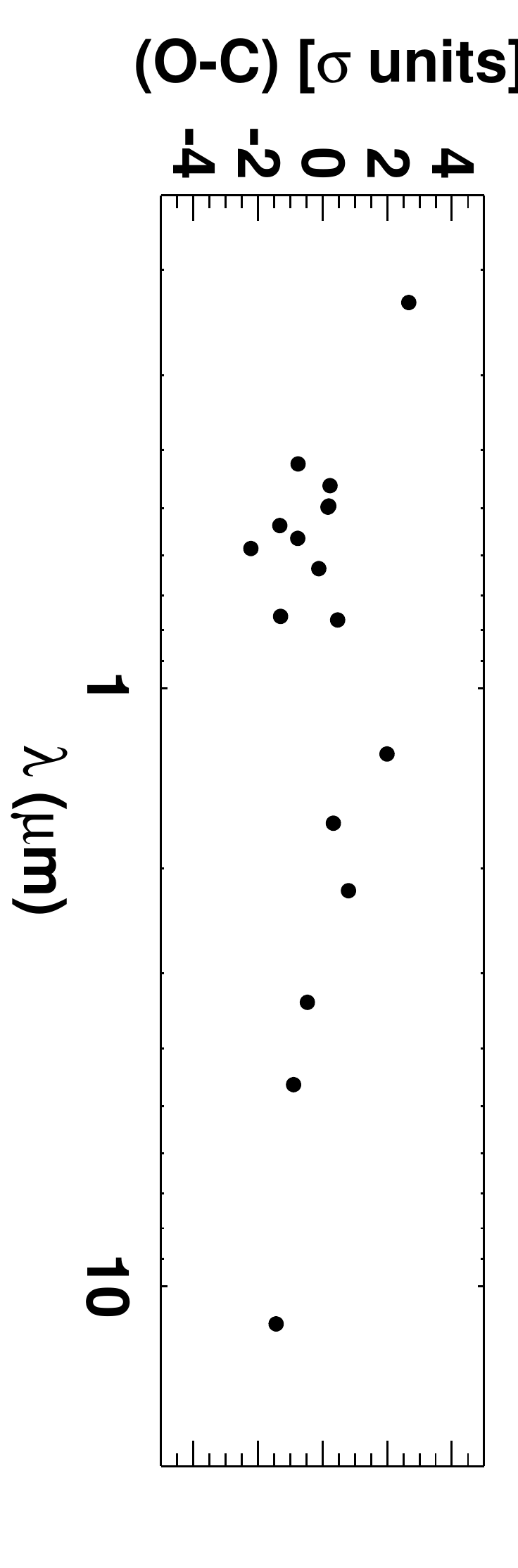}
  \centering\includegraphics[width=6.2cm,angle=90,trim=130 50 50 100,clip]{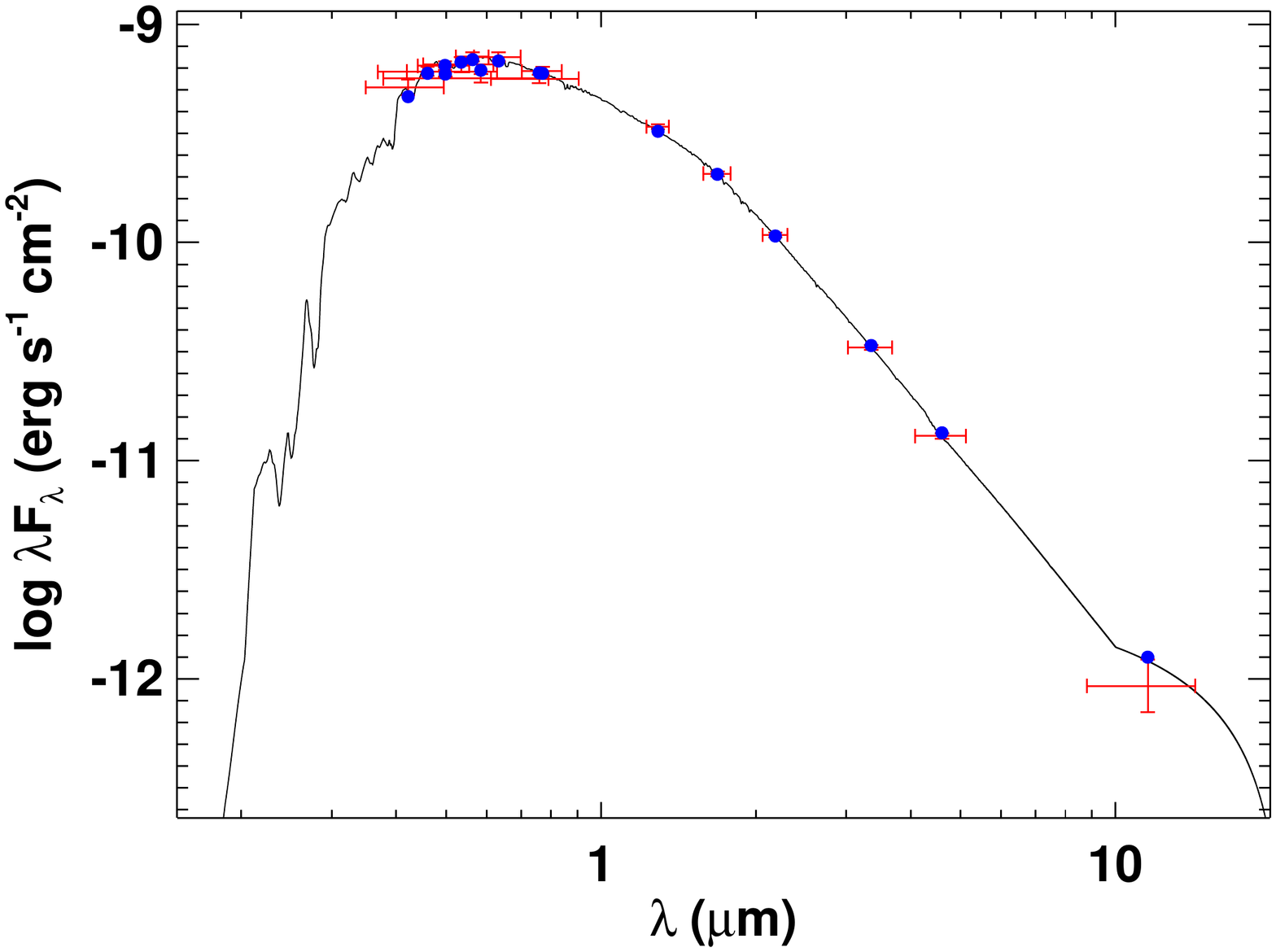}
  \centering\includegraphics[width=3.05cm,angle=90,trim=320 50 80 100,clip]{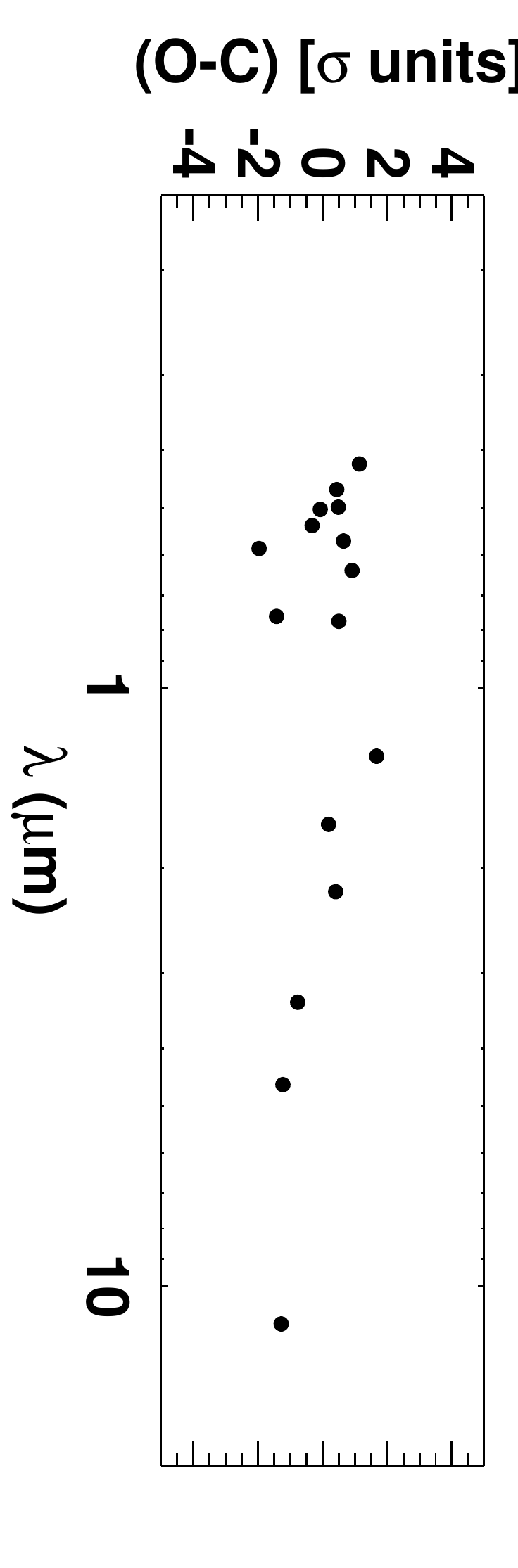}
  \caption{Spectral energy distributions (SEDs) and associated residuals for \ticA\ (top panel) and \ticB\ (bottom panel).
    Red symbols represent the observed photometric measurements, where the horizontal bars represent the effective width of the passband.
    Blue symbols are the model fluxes from the best-fit Kurucz atmosphere model (black).}
  \label{fig:SED_fits}
\end{figure}

\section{Methods}
\label{methods}
\subsection{Stellar parameter determination}
\label{stellar-analysis}

We combined the CORALIE spectra of \ticB\ and the HARPS spectra of \ticA\ in order to derive the parameters for both planet host stars.
The stellar parameters were obtained using the spectral synthesis technique implemented in the iSpec package \citep{blanco-cuaresma_2014}. This package generates a synthetic stellar spectrum using the SPECTRUM radiative transfer code,
the model atmospheres from MARCS \citep{gustafsson_2008}, and the atomic line list from \citet{asplund_2009}.
iSpec minimizes the difference between the observed spectrum and synthetic spectra (computed simultaneously)
and varies only one free parameter at a time.
We defined a set of given wavelength regions for the fitting. The first region includes $H_{\alpha}$, Na, and Mg lines and is used to measure the stellar effective temperature and the surface gravity. The second region includes FeI and FeII lines which are used to derive the stellar metallicity and the projected stellar velocity ($v\sin i$).
We derive the effective temperature, the surface gravity, the metallicity and the $v\sin i$ of \ticA\ and \ticB. We note that both stars are metal-rich with metallicities of $\rm 0.12 \pm 0.08$ and $\rm 0.15 \pm 0.08$ for \ticA\ and \ticB, respectively. The results are presented in Table~\ref{table:stellar-params}.

We performed an analysis of the broadband spectral energy distribution (SED) together with the {\it Gaia\/} EDR3 parallax \citep{gaiacollaboration_2021} in order to determine an empirical measurement of the stellar radius, following the procedures described in \citet{stassun_2016,stassun_2017,stassun_2018}. We pulled the $B_T V_T$ magnitudes from {\it Tycho-2} \citep{hog_2000}, the $BVgri$ magnitudes from APASS \citep{henden_2014}, the $JHK_S$ magnitudes from {\it 2MASS} \citep{skrutskie_2006}, the W1--W4 magnitudes from {\it WISE} \citep{wright_2010}, and the $G$, $G_{\rm BP}$, $G_{\rm RP}$ magnitudes from {\it Gaia}. For \ticA, we also used the available {\it GALEX} NUV flux \citep{bianchi_2017}. Together, the available photometry spans the full stellar SED over the wavelength range 0.35--22~$\mu$m, and extends down to 0.2~$\mu$m for \ticA\ (see Figure~\ref{fig:SED_fits}). We performed a fit using Kurucz stellar atmosphere models, with the priors on effective temperature ($T_{\rm eff}$), surface gravity ($\log g$), and metallicity ([Fe/H]) from the spectroscopically determined values. The remaining free parameter is the extinction ($A_V$), which we restricted to the maximum line-of-sight value from the dust maps of \citet{schlegel_1998}.

Integrating the (dereddened) model SED gives the bolometric flux at Earth of $\rm F_{\rm bol} = 5.86 \pm 0.21 \times 10^{-10} erg\,s^{-1}\,cm^{-2}$ for \ticA\
and $\rm F_{\rm bol} = 8.72 \pm 0.10 \times 10^{-10} erg\,s^{-1}\,cm^{-2}$ for \ticB.
Taking the $F_{\rm bol}$ and $T_{\rm eff}$ together with the {\it Gaia\/} EDR3 parallax
with no systematic adjustment \citep[see][]{stassun_2021} gives the stellar radius, $1.401 \pm 0.045\,R_\odot$ and $1.781 \pm 0.050\,R_\odot$, respectively.
When comparing with the stellar radius estimated from \cite{torres_2010a} empirical relation, we find that the results are identical for \ticA\ but are incompatible for \ticB.
  We find that the spectroscopic log g of \ticB\ may be underestimated:
  part of the spectral line broadening is attributed to rapid rotation instead of gravity broadening.
  A log g value of 4.05 instead of 3.8 allows to obtain compatible stellar radii.

We can also estimate the stellar mass from the empirical relations of \citet{torres_2010a} as well as directly via $R_\star$ and $\log g$.
From \citet{torres_2010a}, the stellar mass is equal to 1.24 $\pm$ 0.07\,$M_\odot$ for \ticA\ and 1.47 $\pm$ 0.09 $M_{\odot}$ for \ticB.
The uncertainty on the stellar masses from the Torres relation is computed by taking into account the uncertainty
  on the stellar parameters as well as the uncertainty on the coefficients of the relation.

Finally, we can estimate the age of the star from the spectroscopic $R'_{\rm HK}$ and from the stellar rotation period determined
from the spectroscopic $v\sin i$ together with $R_\star$, via the empirical relations of \citet{mamajek_2008}.

From \citet{torres_2010a}, the stellar mass is equal to 1.24 $\pm$ 0.07\,$M_\odot$ for \ticA\ and 1.47 $\pm$ 0.09 $M_{\odot}$ for \ticB. The stellar age is estimated from empirical gyrochronology relation \citep{mamajek_2008} and is equal to 5.4 $\pm$ 1.0 Gyr in the case of \ticA\ and in agreement with the age derived from the spectroscopic $R'_{\rm HK}$ equal to 5.7 $\pm$ 1.6 Gyr. However the age estimation is less certain for \ticB\ as the UV activity index points towards a lower stellar rotational velocity than what is derived from the spectroscopic analysis. As a result, the stellar age of \ticB\ ranges from 1.4 to 6.8 Gyr. The results are presented in Table~\ref{table:stellar-params}.

\subsection{Radial velocity analysis}

We first ran an analysis on the radial velocity dataset to find evidence for periodic signal in the data.
We used the \texttt{kima} software package \citep{faria_2018a} to model both radial velocity datasets.
\texttt{kima} uses Bayesian inference to model radial velocity series as a sum of Keplerians,
where different instruments can be included thanks to free radial velocity offset between them.
\texttt{Kima} also allows to have the number of planets as a free parameter.
We chose a uniform prior between 0 and 1 for the number of planets.
The parameters governing the planetary orbit, and the priors used in the fit, are summarized in Table~\ref{table:prior-kima}.

For both systems, there is clear evidence for one periodic signal as the ratio
between the number of posterior samples for the one-planet model over the no planet model is superior to 150.
The orbital period is also clearly defined in both cases: at $20.315 ^{+0.029} _{-0.029}$ days for \ticA\ and $54.32 ^{+0.20} _{-0.16}$ days for \ticB.

The radial velocity time series and their associated generalized Lomb-Scargle periodograms are presented
in Figure~\ref{fig:rv_periodo_1240} for \ticA\ and in Figure~\ref{fig:rv_periodo_2575} for \ticB.

\subsection{Joint analysis}

We perform the joint analysis of the photometric and radial velocity data with the software package \texttt{juliet} \citep{espinoza_2019a}.
\texttt{Juliet} uses Bayesian inference to model a set number of planetary signals.
For the light curve modeling, \texttt{juliet} uses \texttt{batman} \citep{kreidberg_2015} to model the planetary transit,
and the stellar activity as well as instrumental systematics can be taken into account with Gaussian processes \citep{gibson_2014} or simpler parametric functions.
For the radial velocity modeling, \texttt{juliet} uses \texttt{radvel} \citep{fulton_2018a} and stellar activity signal can also be modeled with Gaussian processes.
We chose to use the nested sampling method \texttt{dynesty} \citep{speagle_2019} implemented in \texttt{juliet}.
Several instruments can be taken into account with radial velocity offsets between them.

For \ticA, only two planetary transits have been observed with TESS.
In this case of a duo-transit, the solution in orbital period is a discontinuous space where
several period aliases can explain the two observed transits.
Setting a broad uniform prior on the orbital period leads the algorithm to only explore parts of
the parameter space, usually around one or two of the period aliases. To overcome this difficulty,
we can combine several \texttt{dynesty} runs and thus obtain a more complete picture of the posterior distribution.
We note that the period alias with the highest likelihood corresponds to the orbital period also found with the analysis of the radial velocity data. 

Hence we chose to set a normal prior with a width of 0.1 days on the orbital period
for the joint fit. In order to confirm the correct period alias, we ran the joint modeling with six sets of priors.
We tested the three period aliases closest to the orbital period found with the RV fit (20.1 days). For each period alias prior,
we either fixed the eccentricity to 0 or we let the eccentricity be a free parameter.

For \ticB, the TESS light curves displayed one transit and we obtained two additional transits with NGTS, the solution in orbital period
is thus well constrained and do not present period aliases. For both stars,
the parameters and their prior distributions used for the joint fit are listed in Table~\ref{table:prior-juliet}.
Both targets were analyzed with the same choice of model parameters. For the planet, we have the orbital period, mid-transit time,
impact parameter and planet-to-star radius ratio.
The eccentricity and argument of periastron are used directly as model parameters
and the eccentricity is governed by a Beta prior as detailed in \cite{kipping_2014}. 

We chose to use the stellar density as a parameter
instead of the scaled semi-major axis ($\rm a/R_{\star}$). The normal prior on stellar density is informed by the stellar analysis
which allowed us to derive precise masses and radii and their associated errors.
TESS and NGTS have slightly different bandpasses which we modeled by setting two sets of limb-darkening parameters.
We chose a quadratic limb darkening parameterized as (q1, q2) in order to efficiently sample the parameter space \citep{kipping_2013}.
Additional jitters and offsets are taken into account in the modeling.
As the TESS sectors are two years apart and  NGTS observed the two transits of \ticB\ with different number of cameras,
we have a set of four jitters and offsets for \ticB. For \ticA, we have a set of two jitters and offsets as only two TESS light curves are available.

For the radial velocity model, the semi-amplitude is one of parameters and
each spectrograph has a separate offset and jitter parameters. For both targets, we used the nested sampling algorithm implemented in \texttt{juliet}, \texttt{dynesty}: each fit was done with 1000 live points and until the estimated uncertainty on the log-evidence is smaller than 0.1.

Photometric variability and radial velocity jitter can be accounted for either through linear models against any parameters or through Gaussian processes.
The TESS light curves show small levels of stellar variability both in the SAP and PDCSAP fluxes.
The first NGTS light curve of \ticB\ shows a drop in flux after the end of the first transit as visible in Figure~\ref{fig:phase-folded_lc_2575}.
We tried to decorrelate this feature against relevant parameters extracted from the NGTS observations (e.g. time, airmass, peak flux)
but we were unable to find a combination of linear models which would modeled it.
We chose to model correlated noise in all datasets with a Gaussian process using an approximate Matern kernel implemented via \texttt{celerite} \citep{foreman-mackey_2017}.

\begin{figure}
  \includegraphics[width=\hsize]{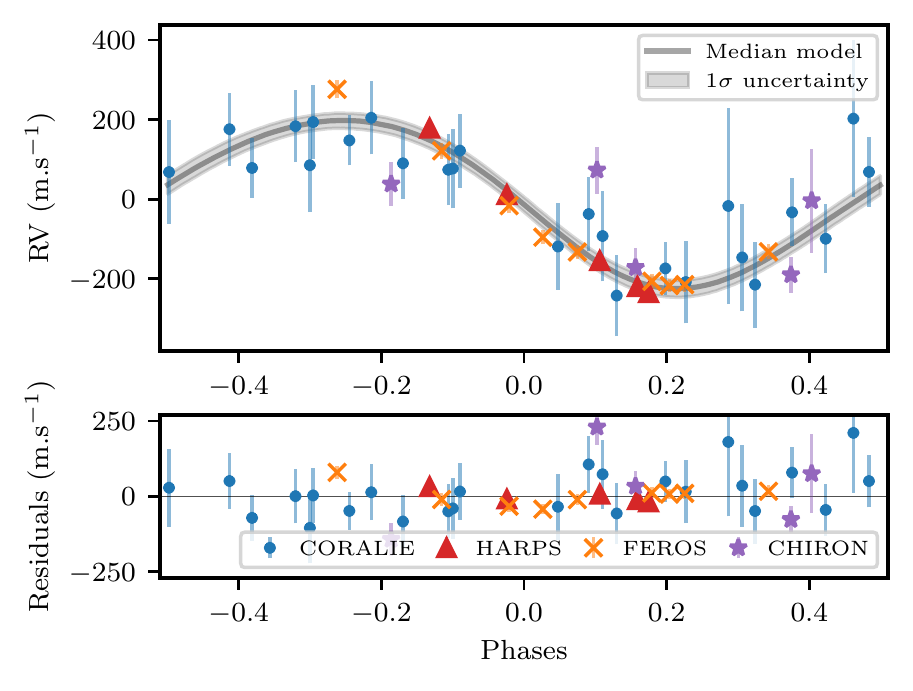}
  \caption{Radial velocities from CORALIE (blue dots), HARPS (red triangles), FEROS (orange crosses), and CHIRON (purple stars) for \ticA. Median Keplerian model is plotted as a grey line along with its corresponding 1 $\sigma$ uncertainty (grey shaded area).}
  \label{fig:rv_timeserie_1240}
\end{figure}

\begin{figure}
  \includegraphics[width=\hsize]{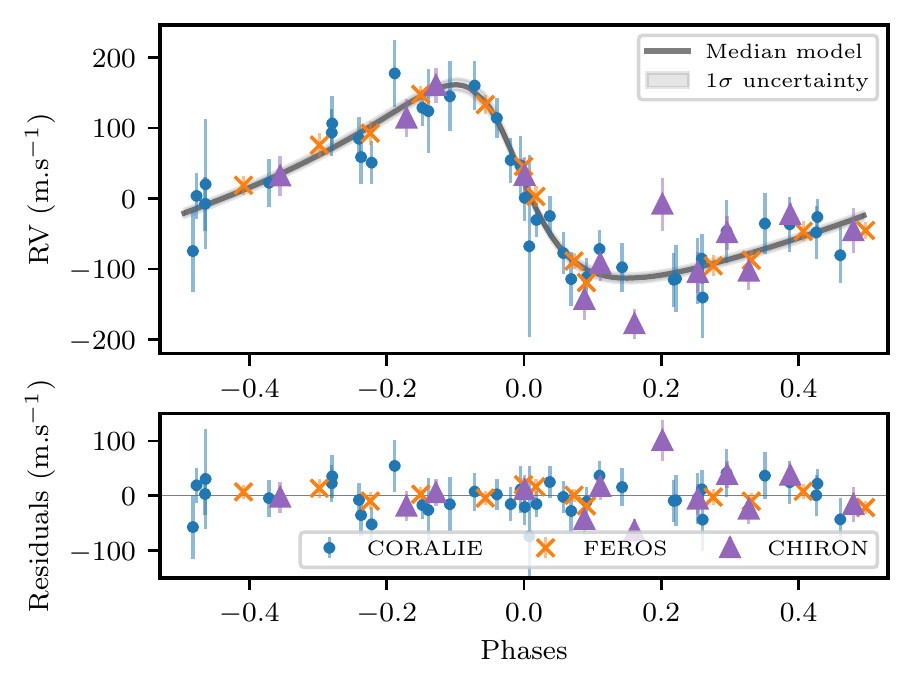}
  \caption{Radial velocities from CORALIE (blue points), FEROS (orange crosses), and CHIRON (purple triangles) for \ticB. Median Keplerian model is plotted as a grey line along with its corresponding 1 $\sigma$ uncertainty (grey shaded area).}
  \label{fig:rv_timeserie_2575}
\end{figure}

\section{Results}
\label{results}

We present the final parameters derived from the joint analysis of \ticA\,b in Section~\ref{results_ticA} and \ticB\,b in Section~\ref{results_ticB}. We compute a first estimate of the heavy element content for both planets in Section~\ref{results_metals}.

\subsection{\ticA}
\label{results_ticA}

For \ticA\ we find that the best solution from the joint fit corresponds to the orbital period of 20.33\,days.
The log-evidence values for the six joint fits are shown in Table~\ref{table:comparison-evidences}.
The highest evidence model favors a non-circular orbit with an eccentricity of $0.091^{+0.024}_{-0.026}$ .
The planet has a radius of $\rm 1.06 ^{+0.04} _{-0.04}\,R_{Jup}$ for a mass of $\rm 3.26 ^{+0.18} _{-0.17}\,M_{Jup}$.
The semi-major axis is equal to $\rm 0.158 ^{+0.006} _{-0.006}\,au$.

Figure~\ref{fig:rv_timeserie_1240} presents the phase folded radial velocities along with the median radial velocity model. The radial velocity semi-amplitude is equal to $\rm 212 ^{+8} _{-8}\,m\,s^{-1}$. The residuals on the radial velocity are about 129, 76, 33, and 15\,$\rm m\,s^{-1}$ for CHIRON, CORALIE, FEROS, and HARPS, respectively. After about one year of radial velocity monitoring of the system, we do not see any hint of a long term drift.

Each TESS sector light curve is modeled with a Gaussian process with an amplitude of 270\,ppm and 560\,ppm and a time-scale of 0.9\,days and 1.4\,days for sectors 6 and 33, respectively. The 30 min binned residuals after model subtraction for the light curve are about 620\,ppm for sector 6 and 490\,ppm for sector 33 .
Phase folded light curves and their corresponding models are shown in Figure~\ref{fig:phase-folded_lc_1240}.
The posterior distributions of the planetary parameters, the stellar density, and the radial velocity semi-amplitude
are presented in Figure~\ref{fig:corner_1240}.
The final parameters of the system can be found in Table~\ref{table:system-parameters}.

\begin{table}
  \caption{Model comparison based on the log-evidence values (ln Z) for \ticA.}
  \label{table:comparison-evidences}
  \centering
  \begin{tabular}{l l l}
    \hline
    \hline
    \noalign{\smallskip}
                           & Free eccentricity       & Fixed eccentricity \\
    \noalign{\smallskip}
    \noalign{\smallskip}
    Period aliases          & ln Z                     & ln Z\\
    \noalign{\smallskip}
    19.78 d                            & 89803.1 $\pm$ 0.5        & 89798.8 $\pm$  0.5   \\
    20.33 d                            & 89815.5 $\pm$ 0.5       & 89808.7 $\pm$  0.5    \\
    20.91 d                            & 89790.2 $\pm$ 0.5      & 89743.0 $\pm$  0.7 \\
    \hline
  \end{tabular}
  \tablefoot{Six fits were performed using a Normal prior on three period aliases around 20 days with the option
    to set the orbital eccentricity as a free or fixed parameter.}
\end{table}

\subsection{\ticB}
\label{results_ticB}

\ticB\ hosts a massive warm Jupiter on a longer period orbit of about 54.19\,days. The planet has a mass of $\rm 2.98 ^{+0.16} _{-0.15}\,M_{Jup}$ and a radius of $\rm 1.07 ^{+0.04} _{-0.04}\,R_{Jup}$. The orbital eccentricity is equal to $0.432 ^{+0.023} _{-0.023}$ and the semi-major axis to $\rm 0.313 ^{+0.0013} _{-0.013}\,au$. 

The radial velocity semi-amplitude is equal to $\rm 138 ^{+5} _{-5}\,m\,s^{-1}$ and the residuals are about 30, 40, and 12\,$\rm m\,s^{-1}$ for CORALIE, CHIRON, and FEROS, respectively. The phase folded radial velocities are presented in Figure~\ref{fig:rv_timeserie_2575}. The radial velocity observations of \ticB\ cover a baseline of two years and there is no hint of a long term drift. 
The three transits are shown in Figure~\ref{fig:phase-folded_lc_2575}. The 30 min binned residuals for the TESS light curve are about 390\,ppm. The two NGTS light curves show residuals, binned to 30 min, close to 525\,ppm for the one camera observation and 130\,ppm for the observation with six cameras.
Figure~\ref{fig:corner_2575} displays the posterior distributions of the planetary parameters along with the stellar density and radial velocity semi-amplitude.
The final parameters of the system are listed in Table~\ref{table:system-parameters}.

\subsection{Heavy element content}
\label{results_metals}

We estimate the heavy element content of both warm Jupiters by comparing their planetary mass and radius
with interior structure models.
We used the planetary evolution model \texttt{completo21} \citep{mordasini_2012} for the core and the envelope modeling
and coupled it with a semi-grey atmospheric model \citep{guillot_2010}.
We select the SCvH equations of state (EOS) of hydrogen and helium (H and He) with a He mass fraction of Y=0.27 \citep{saumon_1995}.
As \cite{thorngren_2016b}, we model the planet with a planetary core of $\rm 10\,M_{\oplus}$.
The core is composed of iron and silicates, with iron mass fraction of 33\%. 
The remaining of heavy elements are homogeneously mixed in the H/He envelope.
The heavy elements are modeled as water with the AQUA2020 EOS of water \citep{haldemann_2020}.

\begin{figure}
  \includegraphics[width=\hsize]{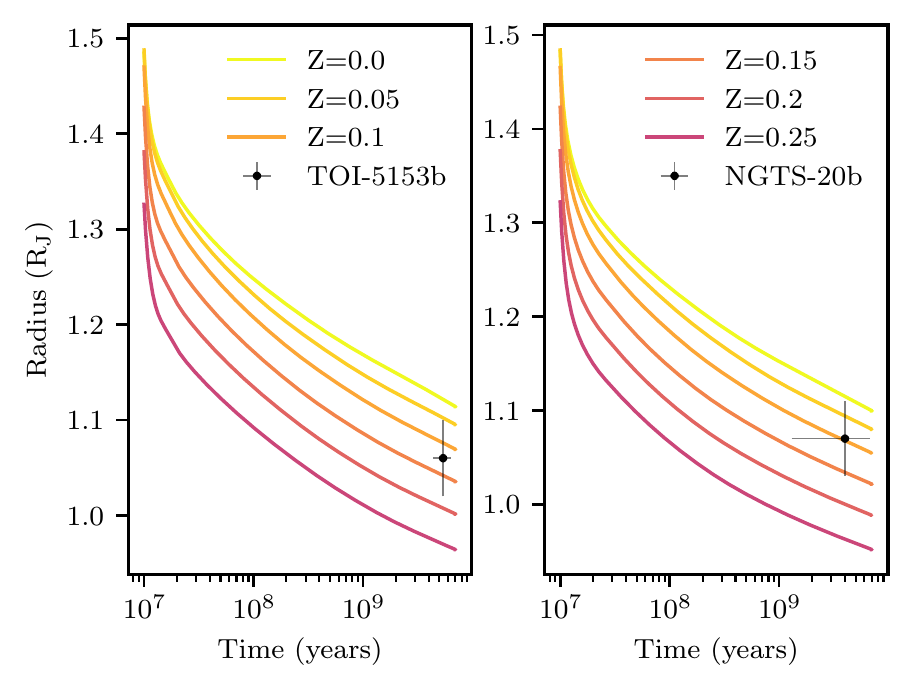}
  \caption{Evolution curves of the planetary radius as a function of time, color-coded by the water mass fraction in the envelope
    for \ticA\,b (left panel) and \ticB\,b (right panel).}
  \label{fig:metal_content}
\end{figure}

We run the evolution tracks for both planets from 10\,Myr to 8\,Gyrs
varying the water mass fraction from 0 to 0.25, as shown in Figure~\ref{fig:metal_content}.
We compute the error on the water mass fraction using a Monte Carlo approach,
taking into account the uncertainties on the planetary radius and the stellar age.
We choose an average value of $\rm 4.1 ^{+2.7} _{-2.7}\, Gyrs$ for the age of \ticB.
We find that the radius of the planets are well explained with a water mass fraction of
$\rm 0.12 ^{+0.06} _{-0.06}$ for \ticA\,b and $\rm 0.10 ^{+0.06} _{-0.06}$ for \ticB\,b
corresponding to a heavy element mass of 133\,$\rm M_{\oplus}$ and 104\,$\rm M_{\oplus}$ respectively.
We vary the planetary mass within the $\rm 1\sigma$ uncertainty and find no significant changes of the water mass fraction.

We assume that the stellar metallicity scales with the iron abundance ([Fe/H]) derived in Section~\ref{stellar-analysis} as follows: 
$\rm Z_{\star} = 0.0142 \times 10^{[Fe/H]}$ \citep{asplund_2009,miller_2011}.
The heavy element enrichment ($\rm Z_{p} / Z_{\star} $) is estimated at $6.9 ^{+3.4} _{-3.4}$ for \ticA\,b and $5.5 ^{+3.1} _{-3.1}$ for \ticB\,b.
We can calculate the heavy element enrichment using \cite{thorngren_2016b} relations and 
we find that $\rm Z_{p} / Z_{\star} $ equals $5.7 ^{+1.0} _{-1.0}$ for \ticA\,b and $5.9 ^{+1.0} _{-1.0}$ for \ticB\,b.
For both planets, our values are in agreement with the estimations derived from \cite{thorngren_2016b}.

\begin{table*}
  \caption{Derived parameters for \ticA\ and \ticB\ systems.}
  \label{table:system-parameters}
  \centering
  \begin{tabular}{l c c}
    \hline
    \hline
    \noalign{\smallskip}
    Parameters                                 & \ticA\,b             & \ticB\,b \\
    \hline
    \noalign{\smallskip}
    Fitted parameters                          &                   & \\
    \noalign{\smallskip}
    Orbital period (days)                     & $20.33003^{+0.00007}_{-0.00007}$              & $54.18915 ^{+0.00015} _{-0.00015}$ \\
    Time of transit $\rm T_0$ (days)          & $2458486.1239^{+0.0019}_{-0.0020}$            & $2458432.9798 ^{+0.0025} _{-0.0025}$\\
    Radius ratio $\rm R_p/R_{\star}$            & $0.0777^{+0.0012}_{-0.0013}$                  & $0.0618 ^{+0.0012} _{-0.0012}$ \\
    Impact parameter                           & $0.725^{+0.024}_{-0.027}$                      & $0.846 ^{+0.014} _{-0.015}$ \\
    Stellar density ($\rm kg\,m^{-3}$)          & $649^{+60}_{-60}$                           & $348 ^{+30} _{-31}$ \\
    TESS limb darkening q1                     & $0.293^{+0.035}_{-0.034}$                    & $0.326 ^{+0.024} _{-0.022}$ \\
    TESS limb darkening q2                     & $0.288^{+0.017}_{-0.017}$                    & $0.298 ^{+0.009} _{-0.009}$  \\
    NGTS limb darkening q1                     & -                                         & $0.380 ^{+0.018} _{-0.019}$  \\
    NGTS limb darkening q2                     & -                                         & $0.330 ^{+0.008} _{-0.009}$ \\
    Eccentricity                               & $0.091^{+0.024}_{-0.026}$                    & $0.432 ^{+0.023} _{-0.023}$  \\
    Argument of periastron (deg)               & $144^{+24}_{-23}$                           & $66.1 ^{+3.2} _{-3.2}$  \\
    Radial velocity semi-amplitude ($\rm m\,s^{-1}$)            & $212 ^{+8} _{-8}$           & $137 ^{+5} _{-5}$    \\
    \noalign{\smallskip}
    \hline
    \noalign{\smallskip}
    Derived parameters                         &                  &\\
    \noalign{\smallskip}
    Planetary radius ($\rm R_{J}$)                &  $1.06 ^{+0.04} _{-0.04}$               & $1.07 ^{+0.04} _{-0.04}$ \\
    Planetary mass ($\rm M_{J}$)                  &  $3.26 ^{+0.18} _{-0.17}$               & $2.98 ^{+0.16} _{-0.15}$ \\
    Inclination (degrees)                      &  $88.27 ^{+0.14} _{-0.14}$                & $88.4 ^{+0.6} _{-0.6}$ \\
    Transit duration (hours)                   &  $4.87 ^{+0.08} _{-0.07}$                 & $4.55 ^{+0.09} _{-0.08}$   \\
    Semi-major axis (au)                       &  $0.158 ^{+0.006} _{-0.006}$               & $0.313 ^{+0.013} _{-0.013}$ \\
     Pericenter distance (au)                  & $0.143 ^{+0.007} _{-0.006}$                & $0.178 ^{+0.011} _{-0.010}$ \\ 
     Apocenter distance (au)                   & $0.172 ^{+0.007} _{-0.007}$                & $0.448 ^{+0.019} _{-0.019}$ \\
     Equilibrium temperature (K)               & $906 ^{+13} _{-13}$                       & $688 ^{+14} _{-13}$ \\
     Equilibrium temperature at periastron (K) & $949 ^{+19} _{-19}$                       & $913 ^{+18} _{-18}$          \\
     Equilibrium temperature at apoastron (K)  & $867 ^{+18} _{-18}$                       & $575 ^{+11} _{-11}$    \\

     \noalign{\smallskip}
    \hline
    \noalign{\smallskip}
    Instrumental parameters                    &                  &\\
    \noalign{\smallskip}
    TESS offset                                & $-0.00005^{+0.00014}_{-0.00012}$       & $-0.00009 ^{+0.00009} _{-0.00009}$   \\
    TESS jitter (ppm)                          & $5^{+26}_{-4}$                       & $2.5 ^{+13.4} _{-2.1}$  \\
    TESS 2 offset                              & $-0.00005^{+0.00028}_{-0.00027}$       & -   \\
    TESS 2 jitter (ppm)                        & $3.6^{+24.3}_{-3.2}$                  & - \\
    NGTS offset                                & -                                  & $0.0002 ^{+0.0017} _{-0.0024}$  \\
    NGTS jitter (ppm)                          & -                                  & $4475 ^{+78} _{-77}$ \\
    NGTS 2 offset                              & -                                  & $-0.0038 ^{+0.0012} _{-0.0006}$\\
    NGTS 2 jitter (ppm)                        & -                                  & $6462 ^{+42} _{-43}$\\
    TESS dilution factor                       & $0.991998 ^{+0.000027} _{-0.000026}$           & -\\

    \noalign{\smallskip}
    GP amplitude TESS  (relative flux)           & $0.00027 ^{+0.00017} _{-0.00007}$      & $0.00027 ^{+0.00008} _{-0.00005}$\\
    GP time-scale TESS  (days)                    & $0.9 ^{+0.7} _{-0.4}$                & $0.61 ^{+0.21} _{-0.14}$ \\
    GP amplitude TESS 2  (relative flux)           & $0.00056 ^{+0.00030} _{-0.00013}$     & - \\
    GP time-scale TESS 2  (days)                   & $1.4 ^{+0.8} _{-0.4}$               & - \\

    GP amplitude NGTS (relative flux)             & -                                & $0.002 ^{+0.006} _{-0.002}$          \\
    GP time-scale NGTS (days)                     & -                                & $4.2 ^{+3.2} _{-2.6}$                \\ 
    GP amplitude NGTS 2 (relative flux)            & -                               & $0.0007 ^{+0.0025} _{-0.0006}$          \\ 
    GP time-scale NGTS 2 (days)                    & -                               & $4.9 ^{+2.9} _{-2.7}$ \\

    \noalign{\smallskip}
    CORALIE offset ($\rm km\,s^{-1}$)            & $-35.396^{+0.018}_{-0.018}$    & $12.553^{+0.005}_{-0.005}$  \\
    CHIRON offset ($\rm km\,s^{-1}$)             & $-36.685 ^{+0.024} _{-0.024}$  & $0.047^{+0.009}_{-0.010}$  \\ 
    FEROS offset ($\rm km\,s^{-1}$)              & $-35.297^{+0.009}_{0.009}$     & $12.585^{0.004}_{0.004}$  \\
    HARPS offset ($\rm km\,s^{-1}$)              & $-35.308^{+0.011}_{-0.012}$    & -  \\ 
    CORALIE jitter ($\rm m\,s^{-1}$)             & $2^{+9}_{-1}$                 & $4^{+4}_{-2}$  \\
    CHIRON jitter ($\rm m\,s^{-1}$)              & $1.1 ^{+3.6} _{-0.9}$          & $26^{+12}_{-10}$  \\
    FEROS jitter ($\rm m\,s^{-1}$)               & $2^{+5}_{-2}$                 & $5^{+5}_{-3}$  \\
    HARPS jitter ($\rm m\,s^{-1}$)               & $0.9^{+3.2}_{-0.7}$            & -  \\
    
  \end{tabular}
\end{table*}

\section{Discussion}
\label{discussion}

The majority of known gas giants with measured masses and radii are hot Jupiters.
They have orbital periods of a few days and are subject to strong interactions with their host star,
such as tidal interactions (e.g. \citealt{valsecchi_2015})
and radius inflation mechanisms (e.g. \citealt{sestovic_2018,sarkis_2021,tilbrook_2021}). There is a positive correlation between the planetary radius and the amount of stellar incident flux \citep{enoch_2012a} and only hot Jupiters receiving a level of stellar irradiation lower than $\rm 2x10^8\,erg\,s^{-1}\,cm^{-2}$ are shown to have a radius independent of the stellar irradiation \citep{demory_2011}.
Cooler gas giants, such as
\ticA\,b and \ticB\,b, with equilibrium temperatures of about 900\,K and 700\,K,
should not be affected by radius inflation mechanisms. Thus we may derive precise bulk metallicities based on evolution models (e.g. \citealt{thorngren_2019a}).
We provide a first estimate of the heavy element content of \ticA\,b and \ticB\,b. We show that their metal-enrichment is in agreement with the mass-metallicity relation and is consistent with planets at long periods with comparable masses (e.g. \citealt{dalba_2022}). Both planets are excellent probes to help us better understand this relation.

We compare the properties of both planets with the population of known transiting exoplanets. We queried the DACE PlanetS exoplanet catalog database\footnote{\href{https://dace.unige.ch/exoplanets/?}{dace.unige.ch/exoplanets}} on February 10th, 2022 and selected exoplanets with mass and radius uncertainties smaller than 25\% and 8\% respectively. The radius uncertainty is scaled to 1/3 of the mass uncertainty to have the same impact on the planetary density.

\begin{figure}
  \includegraphics[width=\hsize]{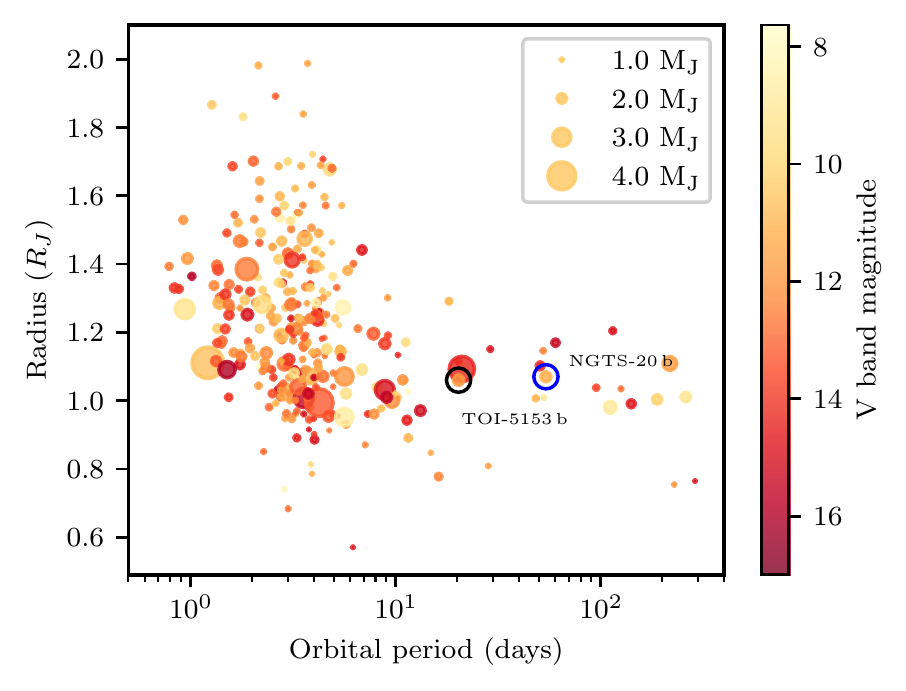}
  \caption{Radius - period diagram for the population of transiting giant planets ($\rm M_{P} > 0.2\,M_{J}$) with mass and radius uncertainties smaller than 25\% and 8\%, respectively.
    The size of the points is proportional to the planetary mass and the V band magnitude is color-coded.
    \ticA\,b is circled in black and \ticB\,b in blue.}
  \label{fig:db_plot}
\end{figure}

Figure~\ref{fig:db_plot} shows that \ticA\,b and \ticB\,b populate a region of the parameter space where fewer systems have been reported with precise mass and radius. Besides, \ticA\,b and \ticB\,b orbit relatively bright stars (Vmag\,=\,11.9 and 11.2) in comparison to the known systems with orbital periods above 20\,days. While the Kepler mission was successful at detecting long-period transiting planets, most of these discoveries were done around faint stars (e.g. \citealt{wang_2015,kawahara_2019}). The follow-up of TESS single transit candidates allows one to probe the same population of planets around brighter stars (e.g. \citealt{eisner_2020,dalba_2022}), which are better suited for follow-up observations. 

\begin{figure}
  \includegraphics[width=\hsize]{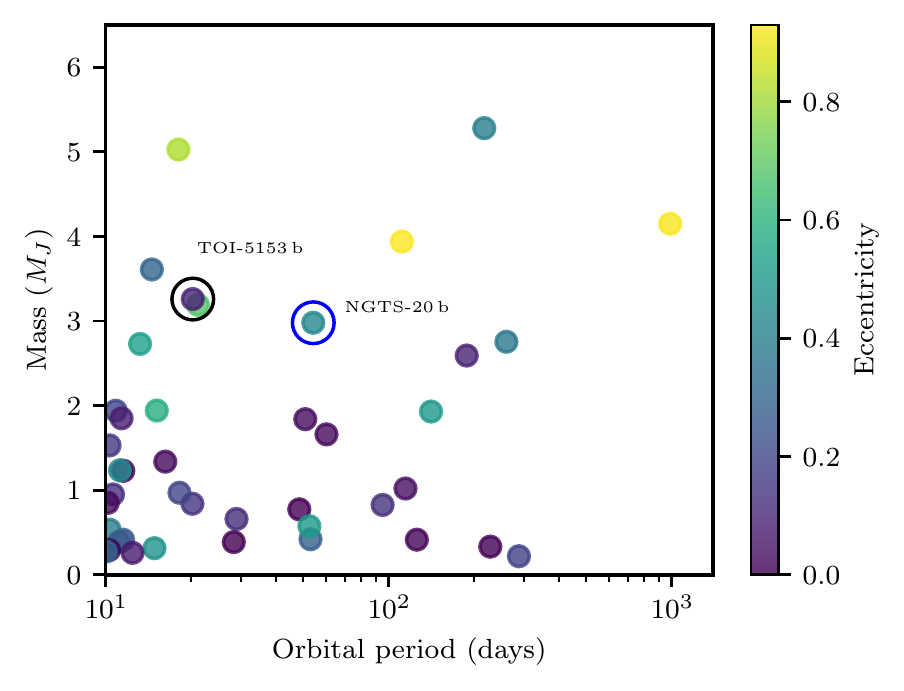}
  \caption{Mass - period diagram for the population of transiting giant planets ($\rm M_{P} > 0.2\,M_{J}$ and $\rm P > 10\,days$) with mass and radius uncertainties smaller than 25\% and 8\% respectively, color-coded as a function of orbital eccentricity. \ticA\,b is circled in black and \ticB\,b in blue.}
  \label{fig:db_plot_2}
\end{figure}

Figure~\ref{fig:db_plot_2} presents the masses, periods, and eccentricities of warm transiting planets.
Despite the small number, we notice that higher mass planets ($M_{P} > 3-4\,M_{J}$) show higher eccentricities.
This trend is reported by \citet{ribas_2007} for a larger sample of planets detected in radial velocities,
where the authors show that the eccentricity distribution of higher mass planets is similar to that of binary stars;
hinting that these planets may have formed by pre-stellar cloud fragmentation.
\ticA\,b and \ticB\,b have similar masses and significantly eccentric orbits with eccentricities of 0.091$\pm$0.026 and 0.43$\pm$0.02, respectively.
\cite{bitsch_2013} showed that the eccentricity of planets with $M_{P} < 5\,M_{J}$ can be damped by the disk.
However, \cite{debras_2021} present disk cavity migration as a possible explanation for eccentricities up to 0.4 for warm Jupiter-mass planets.
Another feature that can be seen is that the highest eccentricities do not occur at closer orbital distances. A possible reason for this high-eccentricity cutoff could be tidal circularization by the host star \citep{adams_2006,dawson_2018}. Planetary orbits beyond a threshold in eccentricity and orbital distance circularize before we observe them \citep{schlecker_2020}. Eccentric warm Jupiters like \ticB\,b can thus serve as a valuable test bed to study tidal interactions between planets and their host stars.

The eccentricities of both targets, and especially \ticB\,b, make them interesting candidates for high-eccentricity migration. Eccentric warm Jupiters could be exoplanets caught in the midst of inward migration. The migration would bring them to close-in orbits, and the orbits of the new hot Jupiters would circularize due to stellar tidal forces. However, other scenarios have been put forward, \cite{schlecker_2020} present the discovery of a warm Jupiter on a highly eccentric 15 day orbit ($\rm e\sim0.58$); the tidal evolution analysis of this system shows that its current architecture likely resulted from an interaction with an undetected companion rather than an on-going high-eccentricity migration.
High-eccentricity migration is a scenario which can be tested by measuring the spin-orbit of the system. Both targets are suitable for Rossiter-McLaughlin observations. The predicted Rossiter-Mclaughlin effect measured with the classical method is about 40\,$\rm m\,s^{-1}$ for \ticA\,b and 16\,$\rm m\,s^{-1}$ for \ticB\,b (Eq 40 from \citealt{winn_2010}). The Rossiter-Mclaughlin effect is large enough so that the spin-orbit angle of the system can be measured with current high-resolution spectrographs.

\section{Conclusions}
\label{conclusion}

We report the discovery of two transiting massive warm Jupiters around the bright and metal-rich stars \ticA\ and \ticB.
\ticA\ hosts a planet on a 20.33\,day period with a planetary mass of 3.26$\pm$0.18\,$M_{J}$ and
planetary radius of 1.06$\pm$0.04\,$R_{J}$. The orbit of the planet has an eccentricity of 0.091$\pm$0.026.
\ticB\ hosts a longer period planet with an orbital period of 54.19\,days.
The planet has a radius of 1.07$\pm$0.04\,$R_{J}$, a planetary mass of 2.98$\pm$0.16\,$M_{J}$,
and presents an eccentric orbit with an eccentricity of 0.43$\pm$0.02.
We show that both planets are metal-enriched and their heavy element content
is consistent with the mass-metallicity relation of gas giants.
We used TESS photometry to identify single transit candidates which were then
followed up with ground-based photometric and spectroscopic instruments
in order to confirm the planetary nature of the transiting objects.
Both warm Jupiters orbit bright stars and are ideal targets
for additional observations in order to measure the spin-orbit alignment of these systems.
These exoplanets show that our selection of targets with single and duo transits and
subsequent radial velocity follow-up are successful and
we expect more discoveries of long-period transiting planets over the coming years.

\begin{acknowledgements}
  This work has been carried out within the framework of the National Centre of
  Competence in Research PlanetS supported by the Swiss National Science
  Foundation under grants 51NF40\_182901 and 51NF40\_205606.
  The authors acknowledge the financial support of the SNSF.

  ML acknowledges support of the Swiss National Science Foundation under grant
  number PCEFP2194576.

  The NGTS facility is operated by the consortium institutes with support from
  the UK Science and Technology Facilities Council (STFC) under projects ST/M001962/1 and ST/S002642/1.
  The contributions at the University of Warwick by PJW, RGW, DRA,
  and SG have been supported by STFC through consolidated grants ST/L000733/1 and ST/P000495/1.
  
  RB acknowledges support from FONDECYT Project 11200751 and from ANID –
  Millennium Science Initiative.

  DD acknowledges support from the TESS Guest Investigator Program grants
  80NSSC21K0108 and 80NSSC22K0185, and NASA Exoplanet Research Program grant
  18-2XRP18\_2-0136.

  Some of the observations in this paper made use of the NN-EXPLORE Exoplanet
  and Stellar Speckle Imager (NESSI). NESSI was funded by the NASA Exoplanet
  Exploration Program and the NASA Ames Research Center. NESSI was built at the
  Ames Research Center by Steve B. Howell, Nic Scott, Elliott P. Horch, and
  Emmett Quigley.

  GZ thanks the support of the ARC DECRA program DE210101893.

  AJ, FR, and PT acknowledge support from ANID -- Millennium  Science  Initiative --
  ICN12\_009 and from FONDECYT project 1210718.

  The results reported herein benefited from collaborations and/or information
  exchange within the program “Alien Earths” (supported by the National
  Aeronautics and Space Administration under agreement No. 80NSSC21K0593) for
  NASA’s Nexus for Exoplanet System Science (NExSS) research coordination
  network sponsored by NASA’s Science Mission Directorate.
  
  JSJ gratefully acknowledges support by FONDECYT grant 1201371 and from the
  ANID BASAL projects ACE210002 and FB210003.

  MNG acknowledges support from the European Space Agency (ESA) as an ESA
  Research Fellow.

  The work performed by HPO has been carried out within the framework of the
  NCCR PlanetS supported by the Swiss National Science Foundation.

  EG gratefully acknowledges support from the David and Claudia Harding
  Foundation in the form of a Winton Exoplanet Fellowship.

  T.T. acknowledges support by the DFG Research Unit FOR 2544
  "Blue Planets around Red Stars" project No. KU 3625/2-1.
  T.T. further acknowledges support by the BNSF program "VIHREN-2021" project No. 
  
  JIV acknowledges support of CONICYT-PFCHA/Doctorado Nacional-21191829.

  The authors acknowledge the use of public TESS data from pipelines at
  the TESS Science Office and at the TESS Science Processing Operations Centre.
  This paper includes data collected with the TESS mission obtained from
  the MAST data archive at the Space Telescope Science Institute (STScI).
  Funding for the TESS mission is provided by the NASA Explorer program.
  STScI is operated by the Association of Universities for Research in Astronomy, Inc.,
  under NASA contract NAS5-26555.

  This work made use of \texttt{tpfplotter} by J. Lillo-Box
  (publicly available in www.github.com/jlillo/tpfplotter),
  which also made use of the python packages \texttt{astropy},
  \texttt{lightkurve}, \texttt{matplotlib} and \texttt{numpy}.
  
  This publication makes use of The Data \& Analysis Center for Exoplanets (DACE),
  which is a facility based at the University of Geneva (CH) dedicated to
  extrasolar planets data visualisation, exchange and analysis.
  DACE is a platform of the Swiss National Centre of Competence in Research (NCCR) PlanetS,
  federating the Swiss expertise in Exoplanet research.
  The DACE platform is available at https://dace.unige.ch.

\end{acknowledgements}

\bibliographystyle{aa}
\bibliography{Photometry-Transit.bib}

\begin{thebibliography}{105}
\expandafter\ifx\csname natexlab\endcsname\relax\def\natexlab#1{#1}\fi

\bibitem[{Adams \& Laughlin(2006)}]{adams_2006}
Adams, F.~C. \& Laughlin, G. 2006, The Astrophysical Journal, 649, 1004

\bibitem[{Albrecht {et~al.}(2012)Albrecht, Winn, Johnson, Howard, Marcy,
  Butler, Arriagada, Crane, Shectman, Thompson, Hirano, Bakos, \&
  Hartman}]{albrecht_2012}
Albrecht, S., Winn, J.~N., Johnson, J.~A., {et~al.} 2012, The Astrophysical
  Journal, 757, 18

\bibitem[{Alibert {et~al.}(2005)Alibert, Mousis, Mordasini, \&
  Benz}]{alibert_2005}
Alibert, Y., Mousis, O., Mordasini, C., \& Benz, W. 2005, The Astrophysical
  Journal, 626, L57

\bibitem[{Aller {et~al.}(2020)Aller, {Lillo-Box}, Jones, Miranda, \&
  Barcel{\'o}~Forteza}]{aller_2020}
Aller, A., {Lillo-Box}, J., Jones, D., Miranda, L.~F., \& Barcel{\'o}~Forteza,
  S. 2020, Astronomy and Astrophysics, 635, A128

\bibitem[{Asplund {et~al.}(2009)Asplund, Grevesse, Sauval, \&
  Scott}]{asplund_2009}
Asplund, M., Grevesse, N., Sauval, A.~J., \& Scott, P. 2009, Annual Review of
  Astronomy and Astrophysics, 47, 481

\bibitem[{Baruteau {et~al.}(2014)Baruteau, Crida, Paardekooper, Masset, Guilet,
  Bitsch, Nelson, Kley, \& Papaloizou}]{baruteau_2014}
Baruteau, C., Crida, A., Paardekooper, S.~J., {et~al.} 2014, Planet-{{Disk
  Interactions}} and {{Early Evolution}} of {{Planetary Systems}} ({eprint:
  arXiv:1312.4293}), 667

\bibitem[{Baxter {et~al.}(2021)Baxter, D{\'e}sert, Tsai, Todorov, Bean, Deming,
  Parmentier, Fortney, Line, Thorngren, Pierrehumbert, Burrows, \&
  Showman}]{baxter_2021}
Baxter, C., D{\'e}sert, J.-M., Tsai, S.-M., {et~al.} 2021, Astronomy \&
  Astrophysics, 648, A127

\bibitem[{Bianchi {et~al.}(2017)Bianchi, Shiao, \& Thilker}]{bianchi_2017}
Bianchi, L., Shiao, B., \& Thilker, D. 2017, The Astrophysical Journal
  Supplement Series, 230, 24

\bibitem[{Bitsch {et~al.}(2013)Bitsch, Crida, Libert, \& Lega}]{bitsch_2013}
Bitsch, B., Crida, A., Libert, A.-S., \& Lega, E. 2013, Astronomy and
  Astrophysics, 555, A124

\bibitem[{{Blanco-Cuaresma} {et~al.}(2014){Blanco-Cuaresma}, Soubiran, Heiter,
  \& Jofr{\'e}}]{blanco-cuaresma_2014}
{Blanco-Cuaresma}, S., Soubiran, C., Heiter, U., \& Jofr{\'e}, P. 2014,
  Astronomy \& Astrophysics, 569, A111

\bibitem[{Brahm {et~al.}(2017)Brahm, Jord{\'a}n, \& Espinoza}]{brahm_2017}
Brahm, R., Jord{\'a}n, A., \& Espinoza, N. 2017, Publications of the
  Astronomical Society of the Pacific, 129, 034002

\bibitem[{Buchhave {et~al.}(2010)Buchhave, Bakos, Hartman, Torres, Kov{\'a}cs,
  Latham, Noyes, Esquerdo, Everett, Howard, Marcy, Fischer, Johnson, Andersen,
  F{\H u}r{\'e}sz, Perumpilly, Sasselov, Stefanik, B{\'e}ky, L{\'a}z{\'a}r,
  Papp, \& S{\'a}ri}]{buchhave_2010}
Buchhave, L.~A., Bakos, G.~{\'A}., Hartman, J.~D., {et~al.} 2010, The
  Astrophysical Journal, 720, 1118

\bibitem[{Buchhave {et~al.}(2014)Buchhave, Bizzarro, Latham, Sasselov, Cochran,
  Endl, Isaacson, Juncher, \& Marcy}]{buchhave_2014}
Buchhave, L.~A., Bizzarro, M., Latham, D.~W., {et~al.} 2014, Nature, 509, 593

\bibitem[{Buchhave {et~al.}(2012)Buchhave, Latham, Johansen, Bizzarro, Torres,
  Rowe, Batalha, Borucki, Brugamyer, Caldwell, Bryson, Ciardi, Cochran, Endl,
  Esquerdo, Ford, Geary, Gilliland, Hansen, Isaacson, Laird, Lucas, Marcy,
  Morse, Robertson, Shporer, Stefanik, Still, \& Quinn}]{buchhave_2012}
Buchhave, L.~A., Latham, D.~W., Johansen, A., {et~al.} 2012, Nature, 486, 375

\bibitem[{Coleman \& Nelson(2016)}]{coleman_2016}
Coleman, G. A.~L. \& Nelson, R.~P. 2016, Monthly Notices of the Royal
  Astronomical Society, 460, 2779

\bibitem[{Cooke {et~al.}(2018)Cooke, Pollacco, West, McCormac, \&
  Wheatley}]{cooke_2018}
Cooke, B.~F., Pollacco, D., West, R., McCormac, J., \& Wheatley, P.~J. 2018,
  Astronomy and Astrophysics, 619, A175

\bibitem[{Dalba {et~al.}(2022)Dalba, Kane, Dragomir, Villanueva, Collins,
  Jacobs, LaCourse, Gagliano, Kristiansen, Omohundro, Schwengeler, Terentev,
  Vanderburg, Fulton, Isaacson, Van~Zandt, Howard, Thorngren, Howell, Batalha,
  Chontos, Crossfield, Dressing, Huber, Petigura, Robertson, Roy, Weiss,
  Behmard, Beard, Brinkman, Giacalone, Hill, Lubin, Mayo, Mo{\v c}nik,
  Akana~Murphy, Polanski, Rice, Rosenthal, Rubenzahl, Scarsdale, Turtelboom,
  Tyler, Benni, Boyce, Esposito, Girardin, Laloum, Lewin, Mann, Marchis,
  Schwarz, Srdoc, Steuer, Sivarani, Unni, Eisner, Fetherolf, Li, Yao, Pepper,
  Ricker, Vanderspek, Latham, Seager, Winn, Jenkins, Burke, Eastman, Lund,
  Rodriguez, Rowden, Ting, \& Villase{\~n}or}]{dalba_2022}
Dalba, P.~A., Kane, S.~R., Dragomir, D., {et~al.} 2022, The Astronomical
  Journal, 163, 61

\bibitem[{Dawson \& Johnson(2018)}]{dawson_2018}
Dawson, R.~I. \& Johnson, J.~A. 2018, Annual Review of Astronomy and
  Astrophysics, 56, 175

\bibitem[{Debras {et~al.}(2021)Debras, Baruteau, \& Donati}]{debras_2021}
Debras, F., Baruteau, C., \& Donati, J.-F. 2021, Monthly Notices of the Royal
  Astronomical Society, 500, 1621

\bibitem[{Demory \& Seager(2011)}]{demory_2011}
Demory, B.-O. \& Seager, S. 2011, The Astrophysical Journal Supplement Series,
  197, 12

\bibitem[{Donati {et~al.}(1997)Donati, Semel, Carter, Rees, \&
  Cameron}]{donati_1997}
Donati, J.-F., Semel, M., Carter, B.~D., Rees, D.~E., \& Cameron, A.~C. 1997,
  Monthly Notices of the Royal Astronomical Society, 291, 658

\bibitem[{Eisner {et~al.}(2020)Eisner, Barrag{\'a}n, Aigrain, Lintott, Miller,
  Zicher, Boyajian, Brice{\~n}o, Bryant, Christiansen, Feinstein,
  {Flor-Torres}, Fridlund, Gandolfi, Gilbert, Guerrero, Jenkins, Jones,
  Kristiansen, Vanderburg, Law, {L{\'o}pez-S{\'a}nchez}, Mann, Safron, Schwamb,
  Stassun, Osborn, Wang, Zic, Ziegler, Barnet, Bean, Bundy, Chetnik, Dawson,
  Garstone, Stenner, Huten, Larish, Melanson, Mitchell, Moore, Peltsch, Rogers,
  Schuster, Smith, Simister, Tanner, Terentev, \& Tsymbal}]{eisner_2020}
Eisner, N.~L., Barrag{\'a}n, O., Aigrain, S., {et~al.} 2020, Monthly Notices of
  the Royal Astronomical Society, 494, 750

\bibitem[{Enoch {et~al.}(2012)Enoch, Collier~Cameron, \& Horne}]{enoch_2012a}
Enoch, B., Collier~Cameron, A., \& Horne, K. 2012, Astronomy \& Astrophysics,
  540, A99

\bibitem[{Espinoza {et~al.}(2019)Espinoza, Kossakowski, \&
  Brahm}]{espinoza_2019a}
Espinoza, N., Kossakowski, D., \& Brahm, R. 2019, Monthly Notices of the Royal
  Astronomical Society, 490, 2262

\bibitem[{Fabrycky \& Tremaine(2007)}]{fabrycky_2007}
Fabrycky, D. \& Tremaine, S. 2007, The Astrophysical Journal, 669, 1298

\bibitem[{Faria {et~al.}(2018)Faria, Santos, Figueira, \& Brewer}]{faria_2018a}
Faria, J.~P., Santos, N.~C., Figueira, P., \& Brewer, B.~J. 2018, The Journal
  of Open Source Software, 3, 487

\bibitem[{{Foreman-Mackey} {et~al.}(2017){Foreman-Mackey}, Agol, Ambikasaran,
  \& Angus}]{foreman-mackey_2017}
{Foreman-Mackey}, D., Agol, E., Ambikasaran, S., \& Angus, R. 2017, The
  Astronomical Journal, 154, 220

\bibitem[{Fulton {et~al.}(2018)Fulton, Petigura, Blunt, \&
  Sinukoff}]{fulton_2018a}
Fulton, B.~J., Petigura, E.~A., Blunt, S., \& Sinukoff, E. 2018, Publications
  of the Astronomical Society of the Pacific, 130, 044504

\bibitem[{F{\"u}r{\'e}sz(2008)}]{furesz_2008}
F{\"u}r{\'e}sz, G. 2008, PhD thesis, University of Szeged

\bibitem[{Gaia~Collaboration {et~al.}(2021)Gaia~Collaboration, Brown,
  Vallenari, Prusti, {de Bruijne}, Babusiaux, Biermann, Creevey, Evans, Eyer,
  Hutton, Jansen, Jordi, Klioner, Lammers, Lindegren, Luri, Mignard, Panem,
  Pourbaix, Randich, Sartoretti, Soubiran, Walton, Arenou, {Bailer-Jones},
  Bastian, Cropper, Drimmel, Katz, Lattanzi, {van Leeuwen}, Bakker, Cacciari,
  Casta{\~n}eda, De~Angeli, Ducourant, Fabricius, Fouesneau, Fr{\'e}mat,
  Guerra, Guerrier, Guiraud, {Jean-Antoine Piccolo}, Masana, Messineo, Mowlavi,
  Nicolas, Nienartowicz, Pailler, Panuzzo, Riclet, Roux, Seabroke, Sordo,
  Tanga, Th{\'e}venin, {Gracia-Abril}, Portell, Teyssier, Altmann, Andrae,
  {Bellas-Velidis}, Benson, Berthier, Blomme, Brugaletta, Burgess, Busso,
  Carry, Cellino, Cheek, Clementini, Damerdji, Davidson, Delchambre, Dell'Oro,
  {Fern{\'a}ndez-Hern{\'a}ndez}, Galluccio, {Garc{\'i}a-Lario},
  {Garcia-Reinaldos}, {Gonz{\'a}lez-N{\'u}{\~n}ez}, Gosset, Haigron, Halbwachs,
  Hambly, Harrison, Hatzidimitriou, Heiter, Hern{\'a}ndez, Hestroffer, Hodgkin,
  Holl, Jan{\ss}en, {Jevardat de Fombelle}, Jordan, {Krone-Martins}, Lanzafame,
  L{\"o}ffler, Lorca, Manteiga, Marchal, Marrese, Moitinho, Mora, Muinonen,
  Osborne, Pancino, Pauwels, Petit, {Recio-Blanco}, Richards, Riello,
  Rimoldini, Robin, Roegiers, Rybizki, Sarro, Siopis, Smith, Sozzetti, Ulla,
  Utrilla, {van Leeuwen}, {van Reeven}, Abbas, Abreu~Aramburu, Accart, Aerts,
  Aguado, Ajaj, Altavilla, {\'A}lvarez, {\'A}lvarez Cid-Fuentes, Alves,
  Anderson, Anglada~Varela, Antoja, Audard, Baines, Baker,
  {Balaguer-N{\'u}{\~n}ez}, Balbinot, Balog, Barache, Barbato, Barros, Barstow,
  Bartolom{\'e}, Bassilana, Bauchet, {Baudesson-Stella}, Becciani, Bellazzini,
  Bernet, Bertone, Bianchi, {Blanco-Cuaresma}, Boch, Bombrun, Bossini,
  Bouquillon, Bragaglia, Bramante, Breedt, Bressan, Brouillet, Bucciarelli,
  Burlacu, Busonero, Butkevich, Buzzi, Caffau, Cancelliere, C{\'a}novas,
  {Cantat-Gaudin}, Carballo, Carlucci, Carnerero, Carrasco, Casamiquela,
  Castellani, {Castro-Ginard}, Castro~Sampol, Chaoul, Charlot, Chemin,
  Chiavassa, Cioni, Comoretto, Cooper, Cornez, Cowell, Crifo, Crosta, Crowley,
  Dafonte, Dapergolas, David, David, {de Laverny}, De~Luise, De~March,
  De~Ridder, {de Souza}, {de Teodoro}, {de Torres}, {del Peloso}, {del Pozo},
  Delbo, Delgado, Delgado, Delisle, Di~Matteo, Diakite, Diener, Distefano,
  Dolding, Eappachen, Edvardsson, Enke, Esquej, Fabre, Fabrizio, Faigler,
  Fedorets, Fernique, Fienga, Figueras, Fouron, Fragkoudi, Fraile, Franke, Gai,
  Garabato, {Garcia-Gutierrez}, {Garc{\'i}a-Torres}, Garofalo, Gavras, Gerlach,
  Geyer, Giacobbe, Gilmore, Girona, Giuffrida, Gomel, Gomez,
  {Gonzalez-Santamaria}, {Gonz{\'a}lez-Vidal}, Granvik,
  {Guti{\'e}rrez-S{\'a}nchez}, Guy, Hauser, Haywood, Helmi, Hidalgo, Hilger,
  H{\l}adczuk, Hobbs, Holland, Huckle, Jasniewicz, Jonker, Juaristi~Campillo,
  Julbe, Karbevska, Kervella, Khanna, Kochoska, Kontizas, Kordopatis, Korn,
  {Kostrzewa-Rutkowska}, Kruszy{\'n}ska, Lambert, Lanza, Lasne, Le~Campion,
  Le~Fustec, Lebreton, Lebzelter, Leccia, Leclerc, {Lecoeur-Taibi}, Liao,
  Licata, Lindstr{\o}m, Lister, Livanou, Lobel, Madrero~Pardo, Managau, Mann,
  Marchant, Marconi, Marcos~Santos, Marinoni, Marocco, Marshall, Martin~Polo,
  {Mart{\'i}n-Fleitas}, Masip, Massari, {Mastrobuono-Battisti}, Mazeh,
  McMillan, Messina, Michalik, Millar, Mints, Molina, Molinaro, Moln{\'a}r,
  Montegriffo, Mor, Morbidelli, Morel, Morris, Mulone, Munoz, Muraveva, Murphy,
  Musella, Noval, Ord{\'e}novic, Orr{\`u}, Osinde, Pagani, Pagano, Palaversa,
  Palicio, Panahi, Pawlak, Pe{\~n}alosa~Esteller, Penttil{\"a}, Piersimoni,
  Pineau, Plachy, Plum, Poggio, Poretti, Poujoulet, Pr{\v s}a, Pulone, Racero,
  Ragaini, Rainer, Raiteri, Rambaux, Ramos, {Ramos-Lerate}, Re~Fiorentin,
  Regibo, Reyl{\'e}, Ripepi, Riva, Rixon, Robichon, Robin, Roelens, Rohrbasser,
  {Romero-G{\'o}mez}, Rowell, Royer, Rybicki, Sadowski,
  Sagrist{\`a}~Sell{\'e}s, Sahlmann, Salgado, Salguero, Samaras,
  Sanchez~Gimenez, Sanna, Santove{\~n}a, Sarasso, Schultheis, Sciacca, Segol,
  Segovia, S{\'e}gransan, Semeux, Shahaf, Siddiqui, Siebert, Siltala, Slezak,
  Smart, Solano, Solitro, Souami, Souchay, Spagna, Spoto, Steele,
  Steidelm{\"u}ller, Stephenson, S{\"u}veges, Szabados, {Szegedi-Elek}, Taris,
  Tauran, Taylor, Teixeira, Thuillot, Tonello, Torra, Torra, Turon, Unger,
  Vaillant, {van Dillen}, Vanel, Vecchiato, Viala, Vicente, Voutsinas, Weiler,
  Wevers, Wyrzykowski, Yoldas, Yvard, Zhao, Zorec, Zucker, Zurbach, \&
  Zwitter}]{gaiacollaboration_2021}
Gaia~Collaboration, G., Brown, A. G.~A., Vallenari, A., {et~al.} 2021,
  Astronomy \& Astrophysics, 649, A1

\bibitem[{Gibson(2014)}]{gibson_2014}
Gibson, N.~P. 2014, Monthly Notices of the Royal Astronomical Society, 445,
  3401

\bibitem[{Gill {et~al.}(2020{\natexlab{a}})Gill, Bayliss, Cooke, Wheatley,
  Nielsen, Lendl, McCormac, Bryant, Acton, Anderson, Belardi, Bouchy, Burleigh,
  Collier~Cameron, Casewell, Chaushev, Goad, G{\"u}nther, Hellier, Jackman,
  Jenkins, Moyano, Pollacco, Raynard, Smith, Tilbrook, Turner, Udry, \&
  West}]{gill_2020}
Gill, S., Bayliss, D., Cooke, B.~F., {et~al.} 2020{\natexlab{a}}, Monthly
  Notices of the Royal Astronomical Society, 491, 1548

\bibitem[{Gill {et~al.}(2022)Gill, {Ulmer-Moll}, Wheatley, Bayliss, Burleigh,
  Acton, Casewell, Watson, Lendl, Worters, Sefako, Anderson, Alves, Bouchy,
  Bryant, Eigm{\"u}ller, Gillen, Goad, Grieves, G{\"u}nther, Henderson,
  Jenkins, Mishra, Moyano, Osborn, Tilbrook, Udry, Vines, \& West}]{gill_2022}
Gill, S., {Ulmer-Moll}, S., Wheatley, P.~J., {et~al.} 2022, {{TIC-320687387
  B}}: A Long-Period Eclipsing {{M-dwarf}} Close to the Hydrogen Burning Limit

\bibitem[{Gill {et~al.}(2020{\natexlab{b}})Gill, Wheatley, Cooke, Jord{\'a}n,
  Nielsen, Bayliss, Anderson, Vines, Lendl, Acton, Armstrong, Bouchy, Brahm,
  Bryant, Burleigh, Casewell, Eigm{\"u}ller, Espinoza, Gillen, R.~Goad,
  Grieves, G{\"u}nther, Henning, Hobson, Hogan, Jenkins, McCormac, Moyano,
  Osborn, Pollacco, Queloz, Rauer, Raynard, Rojas, Sarkis, Smith, Pinto,
  Tilbrook, Udry, Watson, \& West}]{gill_2020a}
Gill, S., Wheatley, P.~J., Cooke, B.~F., {et~al.} 2020{\natexlab{b}}, The
  Astrophysical Journal, 898, L11

\bibitem[{Goldreich \& Tremaine(1980)}]{goldreich_1980}
Goldreich, P. \& Tremaine, S. 1980, The Astrophysical Journal, 241, 425

\bibitem[{Guerrero {et~al.}(2021)Guerrero, Seager, Huang, Vanderburg,
  Garcia~Soto, Mireles, Hesse, Fong, Glidden, Shporer, Latham, Collins, Quinn,
  Burt, Dragomir, Crossfield, Vanderspek, Fausnaugh, Burke, Ricker, Daylan,
  Essack, G{\"u}nther, Osborn, Pepper, Rowden, Sha, Villanueva, Yahalomi, Yu,
  Ballard, Batalha, Berardo, Chontos, Dittmann, Esquerdo, {Mikal-Evans},
  Jayaraman, Krishnamurthy, Louie, Mehrle, Niraula, Rackham, Rodriguez, Rowden,
  {Sousa-Silva}, Watanabe, Wong, Zhan, Zivanovic, Christiansen, Ciardi, Swain,
  Lund, Mullally, Fleming, Rodriguez, Boyd, Quintana, Barclay, Col{\'o}n,
  Rinehart, Schlieder, Clampin, Jenkins, Twicken, Caldwell, Coughlin, Henze,
  Lissauer, Morris, Rose, Smith, Tenenbaum, Ting, Wohler, Bakos, Bean,
  {Berta-Thompson}, Bieryla, Bouma, Buchhave, Butler, Charbonneau, Doty, Ge,
  Holman, Howard, Kaltenegger, Kane, Kjeldsen, Kreidberg, Lin, Minsky, Narita,
  Paegert, P{\'a}l, Palle, Sasselov, Spencer, Sozzetti, Stassun, Torres, Udry,
  \& Winn}]{guerrero_2021}
Guerrero, N.~M., Seager, S., Huang, C.~X., {et~al.} 2021, The Astrophysical
  Journal Supplement Series, 254, 39

\bibitem[{Guillot(2010)}]{guillot_2010}
Guillot, T. 2010, Astronomy \& Astrophysics, 520, A27

\bibitem[{Gustafsson {et~al.}(2008)Gustafsson, Edvardsson, Eriksson,
  J{\o}rgensen, Nordlund, \& Plez}]{gustafsson_2008}
Gustafsson, B., Edvardsson, B., Eriksson, K., {et~al.} 2008, Astronomy and
  Astrophysics, Volume 486, Issue 3, 2008, pp.951-970, 486, 951

\bibitem[{Haldemann {et~al.}(2020)Haldemann, Alibert, Mordasini, \&
  Benz}]{haldemann_2020}
Haldemann, J., Alibert, Y., Mordasini, C., \& Benz, W. 2020, Astronomy \&
  Astrophysics, 643, A105

\bibitem[{Henden \& Munari(2014)}]{henden_2014}
Henden, A. \& Munari, U. 2014, Contributions of the Astronomical Observatory
  Skalnate Pleso, 43, 518

\bibitem[{H{\o}g {et~al.}(2000)H{\o}g, Fabricius, Makarov, Urban, Corbin,
  Wycoff, Bastian, Schwekendiek, \& Wicenec}]{hog_2000}
H{\o}g, E., Fabricius, C., Makarov, V.~V., {et~al.} 2000, Astronomy and
  Astrophysics, v.355, p.L27-L30 (2000), 355, L27

\bibitem[{Howell {et~al.}(2011)Howell, Everett, Sherry, Horch, \&
  Ciardi}]{howell_2011}
Howell, S.~B., Everett, M.~E., Sherry, W., Horch, E., \& Ciardi, D.~R. 2011,
  The Astronomical Journal, 142, 19

\bibitem[{Huang {et~al.}(2020{\natexlab{a}})Huang, Vanderburg, P{\'a}l, Sha,
  Yu, Fong, Fausnaugh, Shporer, Guerrero, Vanderspek, \& Ricker}]{huang_2020}
Huang, C.~X., Vanderburg, A., P{\'a}l, A., {et~al.} 2020{\natexlab{a}},
  Research Notes of the American Astronomical Society, 4, 204

\bibitem[{Huang {et~al.}(2020{\natexlab{b}})Huang, Vanderburg, P{\'a}l, Sha,
  Yu, Fong, Fausnaugh, Shporer, Guerrero, Vanderspek, \& Ricker}]{huang_2020a}
Huang, C.~X., Vanderburg, A., P{\'a}l, A., {et~al.} 2020{\natexlab{b}},
  Research Notes of the American Astronomical Society, 4, 206

\bibitem[{Jenkins {et~al.}(2016)Jenkins, Twicken, McCauliff, Campbell,
  Sanderfer, Lung, {Mansouri-Samani}, Girouard, Tenenbaum, Klaus, Smith,
  Caldwell, Chacon, Henze, Heiges, Latham, Morgan, Swade, Rinehart, \&
  Vanderspek}]{jenkins_2016}
Jenkins, J.~M., Twicken, J.~D., McCauliff, S., {et~al.} 2016, 9913, 99133E

\bibitem[{Jones {et~al.}(2019)Jones, Brahm, Espinoza, Wang, Shporer, Henning,
  Jord{\'a}n, Sarkis, Paredes, {Hodari-Sadiki}, Henry, Cruz, Nielsen, Bouchy,
  Pepe, S{\'e}gransan, Turner, Udry, Marmier, Lovis, Bakos, Osip, Suc, Ziegler,
  Tokovinin, Law, Mann, Relles, Collins, Bayliss, Sedaghati, Latham, Seager,
  Winn, Jenkins, Smith, Davies, Tenenbaum, Dittmann, Vanderburg, Christiansen,
  Haworth, Doty, Fur{\'e}sz, Laughlin, Matthews, Crossfield, Howell, Ciardi,
  Gonzales, Matson, Beichman, \& Schlieder}]{jones_2019}
Jones, M.~I., Brahm, R., Espinoza, N., {et~al.} 2019, Astronomy and
  Astrophysics, 625, A16

\bibitem[{Kaufer {et~al.}(1999)Kaufer, Stahl, Tubbesing, N{\o}rregaard, Avila,
  Francois, Pasquini, \& Pizzella}]{kaufer_1999}
Kaufer, A., Stahl, O., Tubbesing, S., {et~al.} 1999, The Messenger, 95, 8

\bibitem[{Kawahara \& Masuda(2019)}]{kawahara_2019}
Kawahara, H. \& Masuda, K. 2019, The Astronomical Journal, 157, 218

\bibitem[{Kipping(2013)}]{kipping_2013}
Kipping, D.~M. 2013, Monthly Notices of the Royal Astronomical Society, 435,
  2152

\bibitem[{Kipping(2014)}]{kipping_2014}
Kipping, D.~M. 2014, Monthly Notices of the Royal Astronomical Society, 444,
  2263

\bibitem[{Kreidberg(2015)}]{kreidberg_2015}
Kreidberg, L. 2015, Publications of the Astronomical Society of the Pacific,
  127, 1161

\bibitem[{Kurucz(1992)}]{kurucz_1992}
Kurucz, R.~L. 1992, 149, 225

\bibitem[{Lendl {et~al.}(2020)Lendl, Bouchy, Gill, Nielsen, Turner, Stassun,
  Acton, Anderson, Armstrong, Bayliss, Belardi, Bryant, Burleigh, Chaushev,
  Casewell, Cooke, Eigm{\"u}ller, Gillen, Goad, G{\"u}nther, Hagelberg,
  Jenkins, Louden, Marmier, McCormac, Moyano, Pollacco, Raynard, Tilbrook,
  Udry, Vines, West, Wheatley, Ricker, Vanderspek, Latham, Seager, Winn,
  Jenkins, Addison, Brice{\~n}o, Brahm, Caldwell, Doty, Espinoza, Goeke,
  Henning, Jord{\'a}n, Krishnamurthy, Law, Morris, Okumura, Mann, Rodriguez,
  Sarkis, Schlieder, Twicken, Villanueva, Wittenmyer, Wright, \&
  Ziegler}]{lendl_2020}
Lendl, M., Bouchy, F., Gill, S., {et~al.} 2020, Monthly Notices of the Royal
  Astronomical Society, 492, 1761

\bibitem[{Madhusudhan {et~al.}(2014)Madhusudhan, Crouzet, McCullough, Deming,
  \& Hedges}]{madhusudhan_2014a}
Madhusudhan, N., Crouzet, N., McCullough, P.~R., Deming, D., \& Hedges, C.
  2014, The Astrophysical Journal, 791, L9

\bibitem[{Mamajek \& Hillenbrand(2008)}]{mamajek_2008}
Mamajek, E.~E. \& Hillenbrand, L.~A. 2008, The Astrophysical Journal, 687, 1264

\bibitem[{Mayor {et~al.}(2003)Mayor, Pepe, Queloz, Bouchy, Rupprecht, Lo~Curto,
  Avila, Benz, Bertaux, Bonfils, Dall, Dekker, Delabre, Eckert, Fleury,
  Gilliotte, Gojak, Guzman, Kohler, Lizon, Longinotti, Lovis, Megevand,
  Pasquini, Reyes, Sivan, Sosnowska, Soto, Udry, {van Kesteren}, Weber, \&
  Weilenmann}]{mayor_2003a}
Mayor, M., Pepe, F., Queloz, D., {et~al.} 2003, The Messenger, 114, 20

\bibitem[{Melo {et~al.}(2007)Melo, Santos, Gieren, Pietrzynski, Ruiz, Sousa,
  Bouchy, Lovis, Mayor, Pepe, Queloz, {da Silva}, \& Udry}]{melo_2007}
Melo, C., Santos, N.~C., Gieren, W., {et~al.} 2007, Astronomy and Astrophysics,
  Volume 467, Issue 2, May IV 2007, pp.721-727, 467, 721

\bibitem[{Miller \& Fortney(2011)}]{miller_2011}
Miller, N. \& Fortney, J.~J. 2011, The Astrophysical Journal, 736, L29

\bibitem[{Montalto {et~al.}(2020)Montalto, Borsato, Granata, Lacedelli,
  Malavolta, Manthopoulou, Nardiello, Nascimbeni, \& Piotto}]{montalto_2020a}
Montalto, M., Borsato, L., Granata, V., {et~al.} 2020, Monthly Notices of the
  Royal Astronomical Society, 498, 1726

\bibitem[{Mordasini {et~al.}(2012)Mordasini, Alibert, Klahr, \&
  Henning}]{mordasini_2012}
Mordasini, C., Alibert, Y., Klahr, H., \& Henning, T. 2012, Astronomy \&
  Astrophysics, 547, A111

\bibitem[{Mordasini {et~al.}(2014)Mordasini, Klahr, Alibert, Miller, \&
  Henning}]{mordasini_2014}
Mordasini, C., Klahr, H., Alibert, Y., Miller, N., \& Henning, T. 2014,
  Astronomy \& Astrophysics, 566, A141

\bibitem[{Mordasini {et~al.}(2016)Mordasini, {van Boekel}, Molli{\`e}re,
  Henning, \& Benneke}]{mordasini_2016a}
Mordasini, C., {van Boekel}, R., Molli{\`e}re, P., Henning, T., \& Benneke, B.
  2016, The Astrophysical Journal, 832, 41

\bibitem[{Osborn {et~al.}(2016)Osborn, Armstrong, Brown, McCormac, Doyle,
  Louden, Kirk, Spake, Lam, Walker, Faedi, \& Pollacco}]{osborn_2016}
Osborn, H.~P., Armstrong, D.~J., Brown, D. J.~A., {et~al.} 2016, Monthly
  Notices of the Royal Astronomical Society, 457, 2273

\bibitem[{Osborn {et~al.}(2022)Osborn, Bonfanti, Gandolfi, Hedges, Leleu,
  Fortier, Futyan, Gutermann, Maxted, Borsato, Collins, {Gomes da Silva},
  G{\'o}mez Maqueo~Chew, Hooton, Lendl, Parviainen, Salmon, Schanche, Serrano,
  Sousa, Tuson, {Ulmer-Moll}, Van~Grootel, Wells, Wilson, Alibert, Alonso,
  Anglada, Asquier, {Barrado y Navascues}, Baumjohann, Beck, Benz, Biondi,
  Bonfils, Bouchy, Brandeker, Broeg, B{\'a}rczy, Barros, Cabrera, Charnoz,
  Collier~Cameron, Csizmadia, Davies, Deleuil, Delrez, Demory, Ehrenreich,
  Erikson, Fossati, Fridlund, Gillon, {G{\'o}mez-Mu{\~n}oz}, G{\"u}del, Heng,
  Hoyer, Isaak, Kiss, Laskar, {Lecavelier des Etangs}, Lovis, Magrin,
  Malavolta, McCormac, Nascimbeni, Olofsson, Ottensamer, Pagano, Pall{\'e},
  Peter, Piazza, Piotto, Pollacco, Queloz, Ragazzoni, Rando, Rauer, Reimers,
  Ribas, Demangeon, Smith, Sabin, Santos, Scandariato, Schroffenegger, Schwarz,
  Shporer, Simon, Steller, Szab{\'o}, S{\'e}gransan, Thomas, Udry, Walter, \&
  Walton}]{osborn_2022}
Osborn, H.~P., Bonfanti, A., Gandolfi, D., {et~al.} 2022, Uncovering the True
  Periods of the Young Sub-{{Neptunes}} Orbiting {{TOI-2076}}

\bibitem[{Pepe {et~al.}(2002)Pepe, Mayor, Rupprecht, Avila, Ballester, Beckers,
  Benz, Bertaux, Bouchy, Buzzoni, Cavadore, Deiries, Dekker, Delabre,
  D'Odorico, Eckert, Fischer, Fleury, George, Gilliotte, Gojak, Guzman, Koch,
  Kohler, Kotzlowski, Lacroix, Le~Merrer, Lizon, Lo~Curto, Longinotti,
  Megevand, Pasquini, Petitpas, Pichard, Queloz, Reyes, Richaud, Sivan,
  Sosnowska, Soto, Udry, Ureta, {van Kesteren}, Weber, Weilenmann, Wicenec,
  Wieland, {Christensen-Dalsgaard}, Dravins, Hatzes, K{\"u}rster, Paresce, \&
  Penny}]{pepe_2002a}
Pepe, F., Mayor, M., Rupprecht, G., {et~al.} 2002, The Messenger, 110, 9

\bibitem[{Petrovich \& Tremaine(2016)}]{petrovich_2016}
Petrovich, C. \& Tremaine, S. 2016, The Astrophysical Journal, 829, 132

\bibitem[{Petrovich {et~al.}(2014)Petrovich, Tremaine, \&
  Rafikov}]{petrovich_2014}
Petrovich, C., Tremaine, S., \& Rafikov, R. 2014, The Astrophysical Journal,
  786, 101

\bibitem[{Queloz {et~al.}(2001)Queloz, Mayor, Udry, Burnet, Carrier,
  Eggenberger, Naef, Santos, Pepe, Rupprecht, Avila, Baeza, Benz, Bertaux,
  Bouchy, Cavadore, Delabre, Eckert, Fischer, Fleury, Gilliotte, Goyak, Guzman,
  Kohler, Lacroix, Lizon, Megevand, Sivan, Sosnowska, \&
  Weilenmann}]{queloz_2001a}
Queloz, D., Mayor, M., Udry, S., {et~al.} 2001, The Messenger, 105, 1

\bibitem[{Rasio \& Ford(1996)}]{rasio_1996}
Rasio, F.~A. \& Ford, E.~B. 1996, Science, 274, 954

\bibitem[{Ribas \& {Miralda-Escud{\'e}}(2007)}]{ribas_2007}
Ribas, I. \& {Miralda-Escud{\'e}}, J. 2007, Astronomy and Astrophysics, Volume
  464, Issue 2, March III 2007, pp.779-785, 464, 779

\bibitem[{Ricker {et~al.}(2015)Ricker, Winn, Vanderspek, Latham, Bakos, Bean,
  {Berta-Thompson}, Brown, Buchhave, Butler, Butler, Chaplin, Charbonneau,
  {Christensen-Dalsgaard}, Clampin, Deming, Doty, De~Lee, Dressing, Dunham,
  Endl, Fressin, Ge, Henning, Holman, Howard, Ida, Jenkins, Jernigan, Johnson,
  Kaltenegger, Kawai, Kjeldsen, Laughlin, Levine, Lin, Lissauer, MacQueen,
  Marcy, McCullough, Morton, Narita, Paegert, Palle, Pepe, Pepper, Quirrenbach,
  Rinehart, Sasselov, Sato, Seager, Sozzetti, Stassun, Sullivan, Szentgyorgyi,
  Torres, Udry, \& Villasenor}]{ricker_2015a}
Ricker, G.~R., Winn, J.~N., Vanderspek, R., {et~al.} 2015, Journal of
  Astronomical Telescopes, Instruments, and Systems, 1, 014003

\bibitem[{Santerne {et~al.}(2016)Santerne, Moutou, Tsantaki, Bouchy,
  H{\'e}brard, Adibekyan, Almenara, Amard, Barros, Boisse, Bonomo, Bruno,
  Courcol, Deleuil, Demangeon, D{\'i}az, Guillot, Havel, Montagnier,
  Rajpurohit, Rey, \& Santos}]{santerne_2016}
Santerne, A., Moutou, C., Tsantaki, M., {et~al.} 2016, Astronomy and
  Astrophysics, 587, A64

\bibitem[{Sarkis {et~al.}(2021)Sarkis, Mordasini, Henning, Marleau, \&
  Molli{\`e}re}]{sarkis_2021}
Sarkis, P., Mordasini, C., Henning, T., Marleau, G.~D., \& Molli{\`e}re, P.
  2021, Astronomy and Astrophysics, 645, A79

\bibitem[{Saumon {et~al.}(1995)Saumon, Chabrier, \& {van Horn}}]{saumon_1995}
Saumon, D., Chabrier, G., \& {van Horn}, H.~M. 1995, The Astrophysical Journal
  Supplement Series, 99, 713

\bibitem[{Schanche {et~al.}(2022)Schanche, Pozuelos, G{\"u}nther, Wells,
  Burgasser, Chinchilla, Delrez, Ducrot, Garcia, G{\'o}mez Maqueo~Chew,
  Jofr{\'e}, Rackham, Sebastian, Stassun, Stern, Timmermans, Barkaoui,
  Belinski, Benkhaldoun, Benz, Bieryla, Bouchy, Burdanov, Charbonneau,
  Christiansen, Collins, Demory, {D{\'e}vora-Pajares}, {de Wit}, Dragomir,
  Dransfield, Furlan, Ghachoui, Gillon, Gnilka, {G{\'o}mez-Mu{\~n}oz},
  Guerrero, Harris, Heng, Henze, Hesse, Howell, Jehin, Jenkins, Jensen,
  Kunimoto, Latham, Lester, McLeod, Mireles, Murray, Niraula, Pedersen, Queloz,
  Quintana, Ricker, Rudat, Sabin, Safonov, Schroffenegger, Scott, Seager,
  Strakhov, Triaud, Vanderspek, Vezie, \& Winn}]{schanche_2022}
Schanche, N., Pozuelos, F.~J., G{\"u}nther, M.~N., {et~al.} 2022, Astronomy and
  Astrophysics, 657, A45

\bibitem[{Schlecker {et~al.}(2020)Schlecker, Kossakowski, Brahm, Espinoza,
  Henning, Carone, Molaverdikhani, Trifonov, Molli{\`e}re, Hobson, Jord{\'a}n,
  Rojas, Klahr, Sarkis, Bakos, Bhatti, Osip, Suc, Ricker, Vanderspek, Latham,
  Seager, Winn, Jenkins, Vezie, Villase{\~n}or, Rose, Rodriguez, Rodriguez,
  Quinn, \& Shporer}]{schlecker_2020}
Schlecker, M., Kossakowski, D., Brahm, R., {et~al.} 2020, The Astronomical
  Journal, 160, 275

\bibitem[{Schlegel {et~al.}(1998)Schlegel, Finkbeiner, \&
  Davis}]{schlegel_1998}
Schlegel, D.~J., Finkbeiner, D.~P., \& Davis, M. 1998, The Astrophysical
  Journal, 500, 525

\bibitem[{Scott {et~al.}(2018)Scott, Howell, Horch, \& Everett}]{scott_2018}
Scott, N.~J., Howell, S.~B., Horch, E.~P., \& Everett, M.~E. 2018, Publications
  of the Astronomical Society of the Pacific, 130, 054502

\bibitem[{Sestovic {et~al.}(2018)Sestovic, Demory, \& Queloz}]{sestovic_2018}
Sestovic, M., Demory, B.-O., \& Queloz, D. 2018, Astronomy and Astrophysics,
  616, A76

\bibitem[{Showman {et~al.}(2020)Showman, Tan, \& Parmentier}]{showman_2020}
Showman, A.~P., Tan, X., \& Parmentier, V. 2020, Space Science Reviews, 216,
  139

\bibitem[{Sing {et~al.}(2016)Sing, Fortney, Nikolov, Wakeford, Kataria, Evans,
  Aigrain, Ballester, Burrows, Deming, D{\'e}sert, Gibson, Henry, Huitson,
  Knutson, Lecavelier Des~Etangs, Pont, Showman, {Vidal-Madjar}, Williamson, \&
  Wilson}]{sing_2016}
Sing, D.~K., Fortney, J.~J., Nikolov, N., {et~al.} 2016, Nature, 529, 59

\bibitem[{Skrutskie {et~al.}(2006)Skrutskie, Cutri, Stiening, Weinberg,
  Schneider, Carpenter, Beichman, Capps, Chester, Elias, Huchra, Liebert,
  Lonsdale, Monet, Price, Seitzer, Jarrett, Kirkpatrick, Gizis, Howard, Evans,
  Fowler, Fullmer, Hurt, Light, Kopan, Marsh, McCallon, Tam, Van~Dyk, \&
  Wheelock}]{skrutskie_2006}
Skrutskie, M.~F., Cutri, R.~M., Stiening, R., {et~al.} 2006, The Astronomical
  Journal, 131, 1163

\bibitem[{Speagle(2019)}]{speagle_2019}
Speagle, J.~S. 2019, A {{Conceptual Introduction}} to {{Markov Chain Monte
  Carlo Methods}}

\bibitem[{Stassun {et~al.}(2017)Stassun, Collins, \& Gaudi}]{stassun_2017}
Stassun, K.~G., Collins, K.~A., \& Gaudi, B.~S. 2017, The Astronomical Journal,
  153, 136

\bibitem[{Stassun {et~al.}(2018)Stassun, Corsaro, Pepper, \&
  Gaudi}]{stassun_2018}
Stassun, K.~G., Corsaro, E., Pepper, J.~A., \& Gaudi, B.~S. 2018, The
  Astronomical Journal, 155, 22

\bibitem[{Stassun \& Torres(2016)}]{stassun_2016}
Stassun, K.~G. \& Torres, G. 2016, The Astronomical Journal, 152, 180

\bibitem[{Stassun \& Torres(2021)}]{stassun_2021}
Stassun, K.~G. \& Torres, G. 2021, The Astrophysical Journal, 907, L33

\bibitem[{Teske {et~al.}(2019)Teske, Thorngren, Fortney, Hinkel, \&
  Brewer}]{teske_2019}
Teske, J.~K., Thorngren, D., Fortney, J.~J., Hinkel, N., \& Brewer, J.~M. 2019,
  The Astronomical Journal, 158, 239

\bibitem[{Thorngren \& Fortney(2019)}]{thorngren_2019a}
Thorngren, D. \& Fortney, J.~J. 2019, The Astrophysical Journal, 874, L31

\bibitem[{Thorngren {et~al.}(2016)Thorngren, Fortney, {Murray-Clay}, \&
  Lopez}]{thorngren_2016b}
Thorngren, D.~P., Fortney, J.~J., {Murray-Clay}, R.~A., \& Lopez, E.~D. 2016,
  The Astrophysical Journal, 831, 64

\bibitem[{Tilbrook {et~al.}(2021)Tilbrook, Burleigh, Costes, Gill, Nielsen,
  Vines, Queloz, Hodgkin, Worters, Goad, Acton, Henderson, Armstrong, Anderson,
  Bayliss, Bouchy, Briegal, Bryant, Casewell, Chaushev, Cooke, Eigm{\"u}ller,
  Gillen, G{\"u}nther, Hogan, Jenkins, Lendl, McCormac, Moyano, Raynard, Smith,
  Udry, Watson, West, Wheatley, Breytenbach, Sefako, Thomas, \&
  Alves}]{tilbrook_2021}
Tilbrook, R.~H., Burleigh, M.~R., Costes, J.~C., {et~al.} 2021, Monthly Notices
  of the Royal Astronomical Society, 504, 6018

\bibitem[{Tokovinin {et~al.}(2013)Tokovinin, Fischer, Bonati, Giguere, Moore,
  Schwab, Spronck, \& Szymkowiak}]{tokovinin_2013}
Tokovinin, A., Fischer, D.~A., Bonati, M., {et~al.} 2013, Publications of the
  Astronomical Society of the Pacific, 125, 1336

\bibitem[{Torres {et~al.}(2010)Torres, Andersen, \& Gim{\'e}nez}]{torres_2010a}
Torres, G., Andersen, J., \& Gim{\'e}nez, A. 2010, The Astronomy and
  Astrophysics Review, Volume 18, Issue 1-2, pp. 67-126, 18, 67

\bibitem[{Udry {et~al.}(2003)Udry, Mayor, \& Santos}]{udry_2003}
Udry, S., Mayor, M., \& Santos, N.~C. 2003, Astronomy and Astrophysics, 407,
  369

\bibitem[{Valsecchi {et~al.}(2015)Valsecchi, Rappaport, Rasio, Marchant, \&
  Rogers}]{valsecchi_2015}
Valsecchi, F., Rappaport, S., Rasio, F.~A., Marchant, P., \& Rogers, L.~A.
  2015, The Astrophysical Journal, 813, 101

\bibitem[{Villanueva {et~al.}(2019)Villanueva, Dragomir, \&
  Gaudi}]{villanueva_2019a}
Villanueva, Jr., S., Dragomir, D., \& Gaudi, B.~S. 2019, The Astronomical
  Journal, 157, 84

\bibitem[{Wang {et~al.}(2015)Wang, Fischer, Barclay, Picard, Ma, Bowler,
  Schmitt, Boyajian, Jek, LaCourse, Baranec, Riddle, Law, Lintott, Schawinski,
  Simister, Gr{\'e}goire, Babin, Poile, Jacobs, Jebson, Omohundro, Schwengeler,
  Sejpka, Terentev, Gagliano, Paakkonen, Berge, Winarski, Green, Schmitt,
  Kristiansen, \& Hoekstra}]{wang_2015}
Wang, J., Fischer, D.~A., Barclay, T., {et~al.} 2015, The Astrophysical
  Journal, 815, 127

\bibitem[{Wang {et~al.}(2019)Wang, Jones, Shporer, Fulton, Paredes, Trifonov,
  Kossakowski, Eastman, Redfield, G{\"u}nther, Kreidberg, Huang, Millholland,
  Seligman, Fischer, Brahm, Wang, Cruz, Henry, James, Addison, Liang, Davis,
  Tronsgaard, Worku, Brewer, K{\"u}rster, Zhang, Beichman, Bieryla, Brown,
  Christiansen, Ciardi, Collins, Esquerdo, Howard, Isaacson, Latham, Mazeh,
  Petigura, Quinn, Shahaf, Siverd, Rodler, Reffert, Zakhozhay, Ricker,
  Vanderspek, Seager, Winn, Jenkins, Boyd, F{\H u}r{\'e}sz, Henze, Levine,
  Morris, Paegert, Stassun, Ting, Vezie, \& Laughlin}]{wang_2019a}
Wang, S., Jones, M., Shporer, A., {et~al.} 2019, The Astronomical Journal, 157,
  51

\bibitem[{Wheatley {et~al.}(2018)Wheatley, West, Goad, Jenkins, Pollacco,
  Queloz, Rauer, Udry, Watson, Chazelas, Eigm{\"u}ller, Lambert, Genolet,
  McCormac, Walker, Armstrong, Bayliss, Bento, Bouchy, Burleigh, Cabrera,
  Casewell, Chaushev, Chote, Csizmadia, Erikson, Faedi, Foxell, G{\"a}nsicke,
  Gillen, Grange, G{\"u}nther, Hodgkin, Jackman, Jord{\'a}n, Louden,
  Metrailler, Moyano, Nielsen, Osborn, Poppenhaeger, Raddi, Raynard, Smith,
  Soto, \& {Titz-Weider}}]{wheatley_2018}
Wheatley, P.~J., West, R.~G., Goad, M.~R., {et~al.} 2018, Monthly Notices of
  the Royal Astronomical Society, 475, 4476

\bibitem[{Winn(2010)}]{winn_2010}
Winn, J.~N. 2010, arXiv e-prints, 1001, arXiv:1001.2010

\bibitem[{Wittenmyer {et~al.}(2010)Wittenmyer, O'Toole, Jones, Tinney, Butler,
  Carter, \& Bailey}]{wittenmyer_2010}
Wittenmyer, R.~A., O'Toole, S.~J., Jones, H. R.~A., {et~al.} 2010, The
  Astrophysical Journal, 722, 1854

\bibitem[{Wong {et~al.}(2004)Wong, Mahaffy, Atreya, Niemann, \&
  Owen}]{wong_2004}
Wong, M.~H., Mahaffy, P.~R., Atreya, S.~K., Niemann, H.~B., \& Owen, T.~C.
  2004, Icarus, 171, 153

\bibitem[{Wright {et~al.}(2010)Wright, Eisenhardt, Mainzer, Ressler, Cutri,
  Jarrett, Kirkpatrick, Padgett, McMillan, Skrutskie, Stanford, Cohen, Walker,
  Mather, Leisawitz, Gautier, McLean, Benford, Lonsdale, Blain, Mendez, Irace,
  Duval, Liu, Royer, Heinrichsen, Howard, Shannon, Kendall, Walsh, Larsen,
  Cardon, Schick, Schwalm, Abid, Fabinsky, Naes, \& Tsai}]{wright_2010}
Wright, E.~L., Eisenhardt, P. R.~M., Mainzer, A.~K., {et~al.} 2010, The
  Astronomical Journal, 140, 1868

\bibitem[{Wu \& Lithwick(2011)}]{wu_2011}
Wu, Y. \& Lithwick, Y. 2011, The Astrophysical Journal, 735, 109

\bibitem[{Wyttenbach {et~al.}(2017)Wyttenbach, Lovis, Ehrenreich, Bourrier,
  Pino, Allart, {Astudillo-Defru}, Cegla, Heng, Lavie, Melo, Murgas, Santerne,
  S{\'e}gransan, Udry, \& Pepe}]{wyttenbach_2017}
Wyttenbach, A., Lovis, C., Ehrenreich, D., {et~al.} 2017, Astronomy and
  Astrophysics, 602, A36

\end{thebibliography}
\raggedbottom

%
%
\onecolumn
\begin{appendix}

  \section{Radial velocity modeling priors}
  
  \begin{table*}[h!]
    \caption{Priors for the fit of the radial velocities alone.}
  \label{table:prior-kima}
  \centering
  \begin{tabular}{l l l}
    \hline
    \hline
    \noalign{\smallskip}
    Parameters                                 & Distribution     & Value \\
    \hline
    \noalign{\smallskip}
    Number of planets                          & Uniform          & (0, 1)   \\
    Orbital period (days)                      & LogUniform       & (1, 1000)       \\
    Semi-amplitude ($\rm m.s^{-1}$)             & LogUniform       & (1, 1000)       \\ 
    Eccentricity                               & Kumaraswamy      & (0.867, 3.03)       \\
    Mean anomaly (rad)                         & Uniform          & (-$\pi$, $\pi$)       \\
    Argument of periastron (rad)               & Uniform          & (0, 2$\pi$)       \\
    Jitter ($\rm m.s^{-1}$)                     & LogUniform       & (1, 200)       \\
    \hline
  \end{tabular}
  \tablefoot{Priors are identical for both \ticA\ and \ticB.
    The uniform and log uniform distributions are characterized by their lower and upper bounds.}
  \end{table*}

  \section{Joint modeling priors}
  \label{appendix:prior_juliet}

  \begin{table*}[h!]
  \caption{Priors for the joint modeling of photometric and radial velocity data.}
  \label{table:prior-juliet}
  \centering
  \begin{tabular}{l l c c}
    \hline
    \hline
    \noalign{\smallskip}
    Parameters                                 & Distribution     & \ticA & \ticB \\
    \hline
    \noalign{\smallskip}
    Orbital period (days)                      & Normal           & ($P_{alias}$, 0.1) & (54, 0.5) \\
    Time of transit $\rm T_{0}$ (days)         & Uniform          & (2458485.9, 2458486.3) &  (2458432.89,2458433.1)\\
    Radius ratio $\rm R_p/R_{\star}$            & Uniform          & (0, 1)  & (0, 1)  \\
    Impact parameter                           & Uniform          & (0, 1)  &  (0, 1)  \\
    Stellar density ($\rm kg.m^{-3}$)           & Normal           & (685.24, 79.18) & (366.84, 62.71)\\ 
    TESS limb darkening q1                     & Normal           & (0.274, 0.027) & (0.308, 0.038)\\
    TESS limb darkening q2                     & Normal           & (0.275, 0.029) & (0.270, 0.03)\\
    NGTS limb darkening q1                     & Uniform          & - & (0.372, 0.037) \\
    NGTS limb darkening q2                     & Uniform          & - & (0.251, 0.028) \\
    Eccentricity                               & Kumaraswamy / fixed     & (0.867, 3.03) / 0.0  & (0.867, 3.03) \\
    Argument of periastron (deg)               & Uniform          & (0, 360)        & (0, 360)  \\
    TESS offsets                               & Normal           & (0, 0.01)       & (0, 0.01) \\
    TESS jitters (ppm)                         & LogUniform       & (0.1, 1000)     & (0.1, 1000) \\
    NGTS offsets                               & Normal           & -               & (0.0, 0.01) \\
    NGTS jitters (ppm)                         & LogUniform       & -               & (0.1, 10000) \\
    TESS dilution factor                       & Normal           & (0.992, 0.00003)     & -     \\

    \noalign{\smallskip}
    GP amplitude TESS (relative flux)           & LogUniform     & (1e-6, 100.0)  & (1e-6, 100.0)\\
    GP time-scale TESS (days)                   & LogUniform     & (0.001, 100.0)  & (0.001, 100.0)\\
    GP amplitude TESS 2  (relative flux)        & LogUniform     & (1e-6, 100.0)  & - \\
    GP time-scale TESS 2  (days)                & LogUniform     & (0.001, 100.0)  & - \\

    GP amplitude NGTS (relative flux)           & LogUniform     & - & (1e-6, 10.0)   \\
    GP time-scale NGTS (days)                   & Uniform        & - & (0.001, 10.0)   \\ 
    GP amplitude NGTS 2 (relative flux)         & LogUniform     & - & (1e-6, 10.0)   \\ 
    GP time-scale NGTS 2 (days)                 & Uniform        & - & (0.001, 10.0)    \\

    \noalign{\smallskip}
    Semi-amplitude ($\rm km.s^{-1}$)             & Uniform         & (0, 100)        &  (0, 100)    \\
    Spectrograph offsets ($\rm km.s^{-1}$)       & Uniform         & (-100, 100)     & (-100, 100)  \\ 
    Spectrograph jitters ($\rm km.s^{-1}$)       & LogUniform      & (0.001, 0.2)    & (0.001, 0.2)  \\

    \hline
  \end{tabular}
  \tablefoot{Priors differ for \ticA\ and \ticB.
    The normal distribution is defined by two parameters, a mean and a variance.
    For the uniform and log uniform distributions,
    the two parameters define the lower and upper bounds of the distribution.
    $P_{alias}$ can take three values: 20.91, 20.33, 19.78 days and corresponds to period aliases compatible with the radial velocity analysis.
    The priors on the radial velocity offsets and jitters are identical for all spectrographs: CORALIE, CHIRON, FEROS, and HARPS.}
  \end{table*}

  \section{Joint modeling posterior distributions}
  \label{appendix:prior_juliet}

  \begin{figure*}
    \includegraphics[width=0.95\hsize]{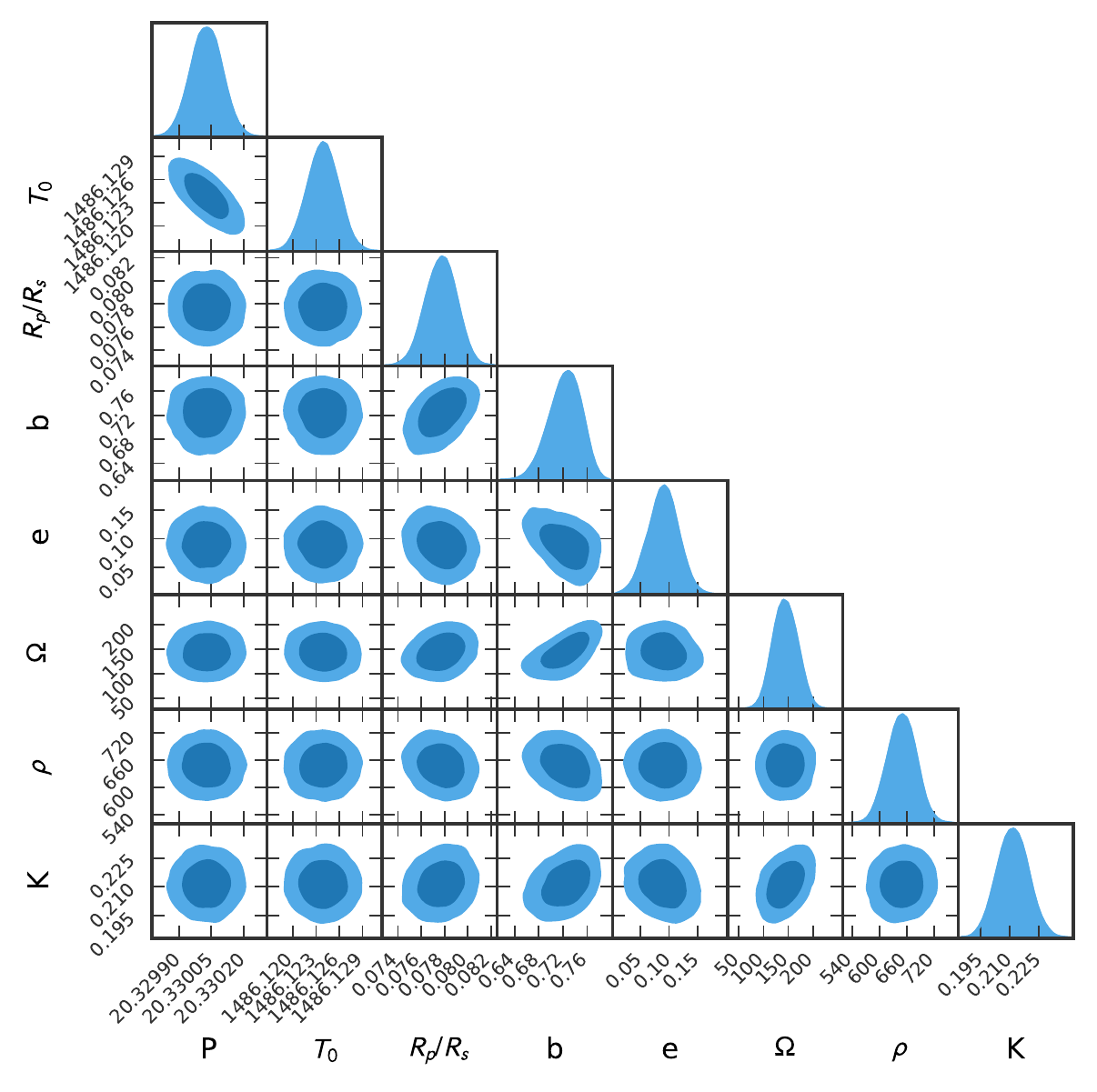}
    \caption{Posterior distributions of fitted parameters for \ticA\,b along with the stellar density ($\rm \rho$) and radial velocity semi-amplitude (K).}
    \label{fig:corner_1240}
  \end{figure*}

  \begin{figure*}
    \includegraphics[width=0.95\hsize]{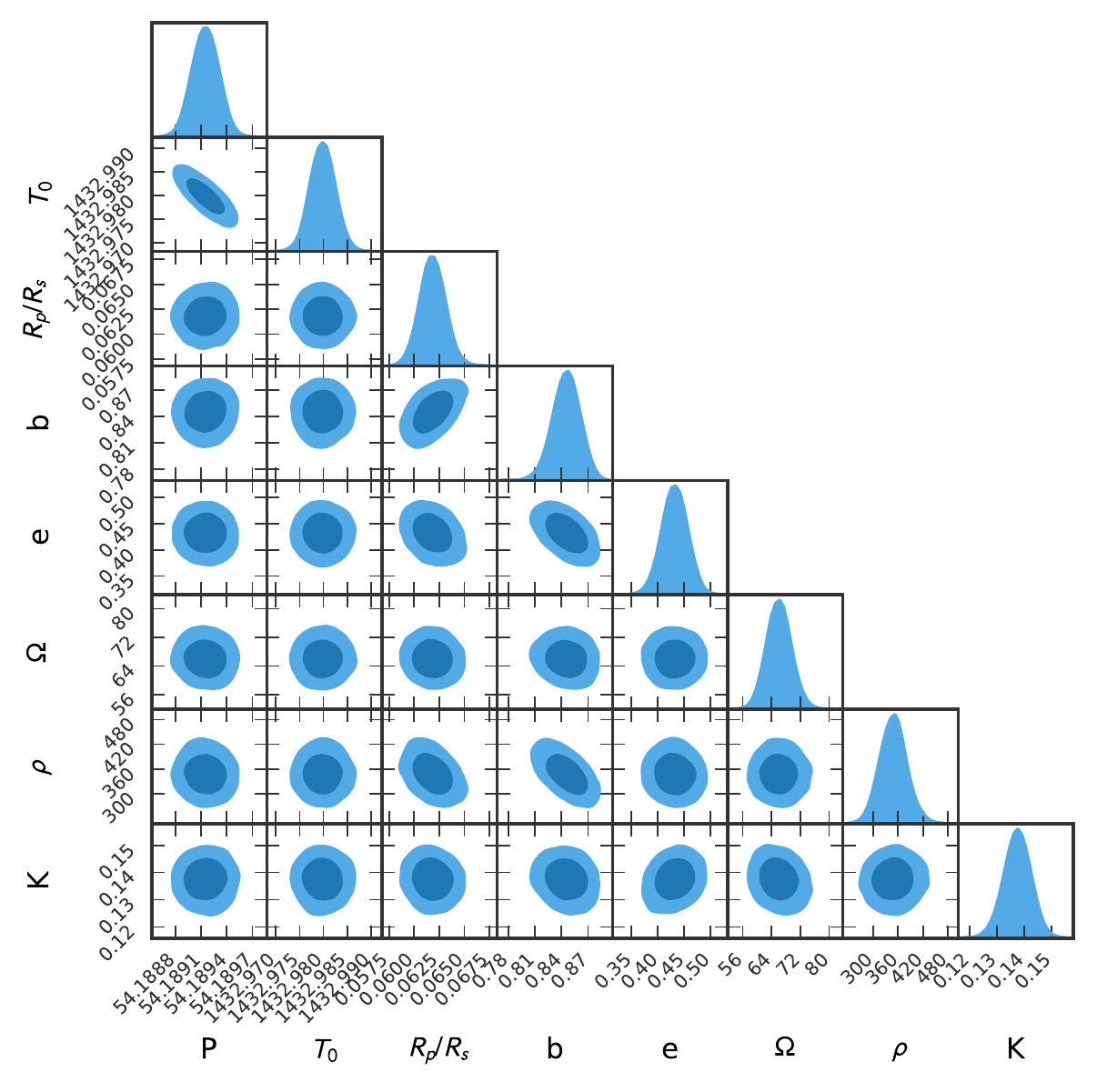}
    \caption{Posterior distributions of fitted parameters for \ticB\,b along with the stellar density ($\rm \rho$) and radial velocity semi-amplitude (K).}
    \label{fig:corner_2575}
  \end{figure*}

\end{appendix}

\vfill
\eject
\end{document}